%% file: paper.tex
\documentclass[sigconf, pageno]{acmart}

\usepackage{subfigure}
\usepackage{enumitem}
\usepackage{pifont}
\usepackage{tikz}
\usepackage{xcolor}
\usepackage[ruled,linesnumbered,vlined]{algorithm2e}
\makeatletter
\newcommand{\removelatexerror}{\let\@latex@error\@gobble}
\makeatother

\AtBeginDocument{%
  }

\setcopyright{acmcopyright}
\copyrightyear{2018}
\acmYear{2018}
\acmDOI{XXXXXXX.XXXXXXX}

\acmConference[e-Energy’24]{The 15th ACM International Conference on Future and Sustainable Energy Systems}{June 4-7, 2024}{Singapore}

\acmPrice{15.00}
\acmISBN{978-1-4503-XXXX-X/18/06}

\definecolor{goldButNotOld}{HTML}{F26035}

\newcommand{\kit}{\texttt{E$^2$-RideKit}\xspace}

\newcommand{\TORA}{\texttt{TORA}\xspace}
\newcommand{\ERA}{\texttt{ERA}\xspace}

\begin{document}

\title[Decarbonizing Ridesharing Platforms]{A Holistic Approach for Equity-aware Carbon Reduction of Ridesharing Platforms}


\author{Mahsa Sahebdel}
\affiliation{%
  \institution{University of Massachusetts Amherst}
  \city{}
  \country{}}

\author{Ali Zeynali}
\affiliation{%
  \institution{University of Massachusetts Amherst}
  \city{}
  \country{}}

\author{Noman Bashir}
\affiliation{%
  \institution{Massachusetts Institute of Technology}
  \city{}
  \country{}}

\author{Prashant Shenoy}
\affiliation{%
  \institution{University of Massachusetts Amherst}
  \city{}
  \country{}}

\author{Mohammad Hajiesmaili}
\affiliation{%
  \institution{University of Massachusetts Amherst}
  \city{}
  \country{}}

\renewcommand{\shortauthors}{Sahebdel et al.}

\begin{abstract}
\input{0-abstract}

\vspace{-0.15cm}
\end{abstract}

%
%




\maketitle

\section{Introduction}
\label{sec:intro}
\input{1-introduction}

\vspace{-2pt}
\section{Problem Statement and Challenges}
\label{sec:problem}

\input{2-problem_challenges}
\section{Ride Assignment Algorithms}
\label{sec:algos}

\input{3-assignment_algorithms}

\section{Route Choice Trade-Offs}
\label{sec:routing_tradeOff}
\input{4-routing}

\section{\kit\ Toolkit}
\label{sec:toolkit}
\input{5-toolkit_description}

\section{Experimental Evaluation}
\label{sec:exp}
\input{6-experiments}

\section{Related Work}
\label{sec:related}
\input{7-related}

\section{Concluding Remarks}
\input{8-conclusion}

\section{Acknowledgements}
\input{9-ack}

\bibliographystyle{ACM-Reference-Format}
\bibliography{paper}

\end{document}

%% file: 0-abstract.tex
In recent years, ridesharing services have revolutionized personal mobility, offering convenient on-demand transportation anytime. While early proponents of ridesharing suggested that these services would reduce the overall carbon emissions of the transportation sector, recent studies reported a type of rebound effect showing substantial carbon emissions of ridesharing platforms, mainly due to their deadhead miles traveled by a ride-share car between two consecutive rides. However, reducing deadhead miles' emissions can incur longer waiting times for riders and starvation of ride assignments for some drivers. Therefore, any efforts towards reducing the carbon emissions from ridesharing platforms must consider the impact on the quality of service, e.g., waiting time, and on the fair and equitable distribution of rides across drivers.

This paper proposes a holistic approach to reduce the carbon emissions of ridesharing platforms while minimizing the degradation in user waiting times and equitable ride assignments across drivers.
Towards this end, we decompose the global carbon reduction problem into two related sub-problems: carbon- and equity-aware ride assignment and fuel-efficient routing. For the ride assignment problem, we consider the trade-off between the amount of carbon reduction and the rider's waiting time and propose simple yet efficient algorithms to handle the conflicting trade-offs.
For the routing problem, we analyze the impact of fuel-efficient routing in reducing the carbon footprint, trip duration, and driver efficiency of ridesharing platforms using route data from Google Maps. Our comprehensive trace-driven experimental results show substantial emissions reduction of our proposed algorithms with only a graceful increase in riders' waiting times. Finally, we release ``\kit'', a toolkit that allows researchers to augment ridesharing datasets with emissions and equity information, enabling further research on emissions analysis and improvement of ridesharing platforms.

%% file: 1-introduction.tex
In 2021, the United States accounted for about 17\% of global greenhouse gas (GHG) emissions. 
Of the 4.6 Billion Metric Tons (BMT) energy-related emissions produced that year, 38\% was from the transportation sector---the highest share among six sectors~\cite{cbo2022emissions}.
Globally, transportation has a similar impact, contributing 37\% of CO\textsubscript{2} emissions in 2021~\cite{iea2022world}. 
In particular, urban mobility is estimated to contribute 40\% of CO\textsubscript{2} emissions from road transport~\cite{eu2023urban}, and given urbanization trends, demand is projected to more than double by 2050~\cite{oke2019novel}. 

The impact of ridesharing services on the transportation market has grown quickly during the past few years as a sustainable alternative to individual vehicle ownership. 
As reported by Bloomberg, the size of ridesharing services was valued at $\$69.3$B in 2022 and is expected to hit $\$205.8$B at the end of 2030~\cite{bloombergRideSharing}. 
These services have revolutionized people's travel by providing convenient access to individual or shared vehicles based on their requested pick-up and drop-off locations. 
Consequently, ridesharing services, such as Uber, Lyft, Grab, and Didi, have become immensely successful due to their promise of personal on-demand mobility at any time~\cite{jiang2018ridesharing}. 

Early proponents of ridesharing suggested that these services would reduce reliance on privately-owned cars, reduce traffic congestion, and reduce carbon emissions, with early studies estimating that at least five private vehicles would be replaced for each shared car and there would likely be carbon emission reductions if shared cars were newer vehicles~\cite{upenn}.
However, the success of these services has resulted in an increase in traffic and more congestion on roads---a rebound effect~\cite{Henao2017}. For example, in New York City, ridesharing has been shown to constitute more than 50\% of road traffic~\cite{citylab,Schaller2017}.
Another study has estimated that a typical ridesharing trip is less efficient than a personal car trip, mainly due to ``deadhead'' miles traveled by a ride-share vehicle between consecutive hired rides, and that this generates 47\% more CO$_2$ emissions than an equivalent private car ride~\cite{ucsusa,reuters}. 
The study also showed that the greater convenience of ridesharing has steered passengers away from public transit options. 
Others have reported that the deadhead miles of ridesharing services account for an estimated 36-45\% of their overall distance traveled~\cite{komanduri2018assessing, henao2019impact, cramer2016disruptive}. 

The prevalence of deadhead miles and resulting carbon emissions presents new opportunities to develop emission-aware ride assignment algorithms. 
However, any carbon-aware optimization of ridesharing systems must consider its impact on riders and drivers.
Prior work shows that existing ride assignment algorithms may be unfair to drivers from certain demographics based on their gender, age, and race~\cite{rosenblat2016discriminating, cook2021gender, graham2017towards}.
An emission-aware ride assignment can perpetuate such disparities by prioritizing drivers with newer, more fuel-efficient vehicles.  At the same time, new emission-aware ride assignment algorithms should not increase rider's waiting times, which can negatively impact user satisfaction.  
Therefore, any efforts towards reducing the carbon emissions from ridesharing platforms should carefully consider its impact on the quality of service (QoS), e.g., waiting time, and equity, e.g., distribution of rides across drivers. While there has been significant work on optimizing various facets of ridesharing systems \cite{fielbaum2022optimizing, wang2023optimization, mourad2019survey}, the problem of emission-aware ride assignment to green ridesharing systems has not seen much research attention.

The design space for emission-aware ridesharing optimization is two-fold:  (1) ride assignment to reduce the emissions from the deadhead miles between two trips; (2) the routing strategy of a trip to optimize emissions instead of other factors such as trip duration or distance. In both problems, there is a trade-off between emissions, QoS, and equity implications.
A ride assignment based on a single objective of emissions reduction may degrade both the \textit{rider} and \textit{driver}'s QoS. 
For example, emission-aware ride assignments may increase the rider's waiting time by picking a vehicle from a soon-to-be-completed trip to minimize the deadhead miles. 
Similarly, during trip routing, the extended time in picking the most fuel-efficient route instead of the fastest route may degrade user waiting times. 
The equity impacts of emission-aware ride assignments are more subtle. 
The emission-aware ride assignment may prefer electric or low-emissions vehicles, which is unfair to drivers with high-emissions vehicles that tend to belong to low-income communities. 
Alternatively, the emissions-aware ride assignment policy may assign trips with inevitably higher deadhead miles to low-emissions vehicles, which then decreases the efficiency of their service due to a higher deadhead-to-trip ratio. 

In this paper, we take a holistic approach to emission-aware ride assignment and trip routing in ridesharing platforms that considers both the QoS and equity implications.
We analyze the performance and the trade-offs between their emissions reduction and QoS of riders and drivers using a real-world ridesharing dataset. 
In doing so, our paper makes the following contributions: 

\vspace{-0.1cm}
\begin{enumerate}[leftmargin=*, itemsep=0.05cm]
    \item \textbf{Emission-aware ride assignment:} 
    We develop emissions-aware ride assignment algorithms aimed at minimizing emissions from deadhead miles while maintaining the quality of service (QoS) for both riders and drivers. We first introduce an offline solution that showcases the substantial potential for emission reduction through an emission-aware ride assignment policy. Following that, we introduce an online strategy that allows for a configurable trade-off between emission reductions and QoS, as measured by rider waiting times.

    \item \textbf{Trip routing:} In addition to the ride assignment strategy, the routing policy also plays a significant role in determining overall carbon emissions. We utilize route data provided by popular navigation apps such as Google Maps to evaluate the trade-offs among time (the fastest route), distance (the shortest route), and fuel consumption (the greenest route) in routing following emission-aware ride assignments.
    

    \item \textbf{\kit design:} As an independent contribution, we present \kit, which augments an arbitrary ridesharing dataset with per-ride carbon emission and per-driver/passenger socioeconomic demographic information. It allows users to configure EV penetration in the dataset, plugin complementary models, and extend the toolkit to add additional information. As an example use case, we augment the widely-used RideAustin dataset~\cite{rideaustin-dataset}  with carbon emission and equity information. 
    In doing so, we outline the challenges we faced in augmenting the dataset and how we solved them. We hypothesize that \kit enables researchers to augment ridesharing datasets with emissions and equity information, enabling new ridesharing analytics and optimizations. 
    
    
    \item\textbf{Experimental results:} Finally, leveraging our toolkit, we conduct a comprehensive evaluation of our emission-aware ride assignment algorithm and trip routing policies using the RideAustin dataset to produce multiple key findings.  \textbf{First}, we demonstrate that our emission-aware ride assignment algorithms can decrease emissions from deadhead miles by up to 60\% with a mere 4\% increase in waiting time for the riders. \textbf{Second}, replacing just 5\% of the current fleet with EVs can improve deadhead miles emission reduction to 67\%. \textbf{Third}, optimizing trip routing for emissions does not result in favorable tradeoffs; emissions can only be reduced by 4.2\% w.r.t. the fastest route while increasing the trip duration by 3\%. \textbf{Finally}, combining emission-aware ride assignment and trip routing can reduce the fleet-level emissions by 26\%, at the expense of a 6.2\% increase in trip duration. The emission reduction can be improved to 29\% by replacing only 5\% of the current fleet with EVs.

\end{enumerate}

%% file: 2-problem_challenges.tex
Addressing the general problem of emissions reduction in ridesharing\footnote{Our work does not address carpooling or ridesharing of passengers.}platforms involves two main challenges. 
First, the routing of trips within the platform is important. Routing algorithms can be tailored not only for the shortest distance or fastest route but also to minimize fuel consumption and emissions. 
Second, optimizing ride assignments to reduce emissions is also crucial. This involves efficiently allocating vehicle rides to minimize deadhead mile between the current trip's drop-off location and the next pick-up location. Achieving this goal is complex because it requires balancing reducing emissions and maintaining a high-quality experience for drivers and riders. 
These two interconnected components of the problem require careful consideration and trade-offs to ensure a sustainable and equitable ridesharing ecosystem that simultaneously addresses environmental and service quality concerns.

\noindent\textit{The fuel-efficient routing problem.}
Routing algorithms are pivotal in determining the optimal route during a trip. These algorithms consider various factors, including travel time, distance, and emissions produced. Variables such as route distance, traffic conditions, and road quality influence travel time and emissions associated with a particular route. It is important to note that the route with the lowest emissions may not always align with the shortest distance or the fastest travel time. As a result, different routing algorithms may yield varying emissions and waiting times results.

\noindent\textit{The emission-aware ride assignment.}
The ride assignment strategy also impacts the emissions due to the deadhead mile of a trip.
Since drivers are located at varying distances from passengers, and their vehicles have distinct unit-distance emissions (emissions produced per unit of distance traveled), the passenger-to-driver assignment algorithm significantly impacts the overall emissions of the ridesharing system and the average waiting time for passengers.
While some simplistic assignment algorithms may reduce emissions for individual trips by assigning passengers to drivers with the lowest emissions, this approach can lead to substantially extended waiting times if the low-emission driver is situated far from the passenger, potentially degrading the passenger's satisfaction.
Furthermore, assigning passengers to drivers with the lowest emissions does not inherently result in long-term optimal ride assignments for the entire ridesharing service. On the other hand, a strategy that prioritizes assigning passengers to the nearest available drivers aims to minimize waiting times but may inadvertently result in higher carbon emissions, as it overlooks variations in unit-distance emissions produced by different vehicles in the fleet.

\begin{figure}
    \centering
    \includegraphics[width=0.7\linewidth]{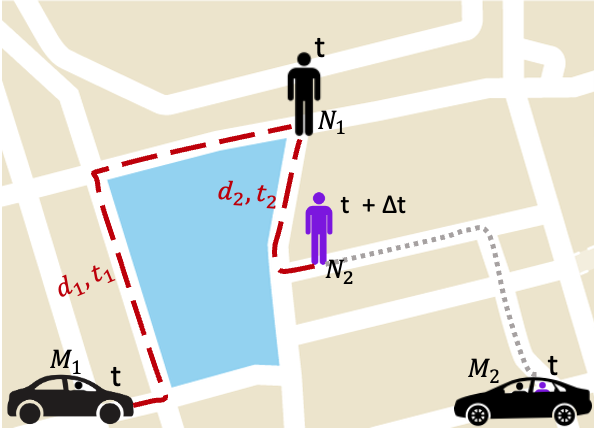}
    \vspace{-0.25cm}
    \caption{\emph{Ride assignment with the goal of minimizing deadhead miles yields lower such miles and increment in waiting time. We can reduce deadhead miles by assigning passenger $N_1$ to driver $M_2$ with the cost of increase in waiting time.}}
    \vspace{-10pt}
    \label{fig:illustration}
\end{figure}

\noindent\textit{An illustrative example.}
In Figure~\ref{fig:illustration}, we illustrate the benefits of our system-level emissions-aware ride assignment over per-ride optimization strategies~\cite{kontou2020reducing}. 
Consider a scenario where a passenger, identified as $N_1$, requests a ride at time $t$. At that moment, a driver, labeled as $M_1$, is available for assignment with associated deadhead miles denoted as $d_1$. Concurrently, another driver, $M_2$, has already been assigned to passenger $N_2$ at time $t$. Now, if we consider a situation where passenger $N_1$ is willing to wait for a duration of $\Delta t$ until being assigned to $M_2$, it leads to a reduction in deadhead miles, now represented as $d_2 < d_1$. However, this reduction in deadhead miles comes at the expense of an increase in the passenger's waiting time, shifting from $t_1$ to $\Delta t + t_2$.

\noindent\textit{Problem formulation.}
At a high level, the ride-assignment problem is an extended version of the classic online bipartite matching problem where two sets of disjoint nodes represent riders and drivers. However, modeling the topology constraints, i.e., determining the set of available drivers and their corresponding assignment cost for a new ride request, is highly challenging since other rides might already occupy some drivers, where some of those drivers might be available soon since they are close to finishing their current ride. A rigorous formulation must carefully capture multiple complex topology-constrained and time-coupled relationships between riders and drivers. 
We formulate the emission-aware one-on-one ride assignments within a ridesharing platform by considering a high-level abstraction of the topology constraint of the problem in an online setting. We consider a ridesharing platform comprising $M$ drivers. Over a certain time horizon, $N$ ride requests are generated. The objective of the ride assignment algorithm is to assign an available empty car to each request such that it minimizes the long-term emissions from the total distance traveled. This encompasses rides with passengers and the deadhead mile between dropping off one passenger and picking up the next. We describe a simplified version of the Emission-aware Ride Assignment Problem (\texttt{ERAP}) as follows.
\begin{align*}
    [\texttt{ERAP}]\quad \min &\sum_{n=1}^N \sum_{m=1}^{M_n} \left(e_t(n,m) + e_d(n,m)\right) x_{n,m} \\
    \text{s.t.,} & \sum_{m=1}^{M_n} x_{n,m} = 1, \quad n \in [N], \\
    \text{vars.,} & \quad x_{n,m} = \{0,1\}, \quad n \in [N], m \in [M_n],
\end{align*}
where $e_t(n,m)$ are the trip emissions due to the assignment of passenger $n$ to driver $m$ and $e_d(n,m)$ represents the emissions due to the deadhead mile for driver $m$ to pickup passenger $n$. 
The routing algorithms significantly impact these two terms. Also, $x_{n,m}$ is derived from a ride assignment algorithm and is a binary optimization variable where $x_{n,m}=1$ if $m$ is assigned to $n$; $0$, otherwise.

\noindent\textit{Balancing the maximum potential of emissions reduction and riders' QoS.}
In \texttt{ERAP}, parameter $M_n$ denotes the set of available cars for passenger $n$ that could include both currently available and soon-to-be-completed cars for ride $n$. Hence, $M_n$ is a crucial parameter that balances the maximum potential of emissions reduction and riders' waiting time (QoS), i.e., the more willingness to increase the waiting time of ride $n$, the more cars will be eligible to be included in $M_n$. In our algorithm in Section~\ref{sec:tora}, we define a threshold parameter $\phi$ that determines the set of available cars for a new ride, and by adjusting parameter $\phi$, the set of available cars $M_n$ changes; hence, the algorithm could be tuned to achieve a desired trade-offs between emission reduction and user waiting time.  

Lastly, we note that one could implicitly leverage parameter $M_n$ to impose driver's equity constraints. That is, there could be another separate module that records the equity-related metrics for drivers, e.g., deadhead-to-trip ratio and the number of assigned trips, and then based on the equity status of each driver, the available cars for the new ride might be determined, e.g., the driver equity module may exclude the drivers with a large number of previously assigned rides to make the ride assignment equitable for other available drivers. That said, we emphasize that our current modeling does not \textit{explicitly} model the optimization of equity from the drivers' perspective. Designing equitable ride assignments for drivers is a significant further direction of our work.

%% file: 3-assignment_algorithms.tex
In Section~\ref{sec:era}, we first present an offline ride assignment algorithm, \ERA, designed to assign rides to minimize total emissions near optimally. While \ERA is not a practical algorithm, it can show the maximum potential of emissions reduction with the sole objective of reducing the emissions. In Section~\ref{sec:tora}, we then present \TORA, an online emission-aware ride assignment algorithm that balances total emissions and the average passenger waiting time. 

\subsection{\ERA: An Offline Emission-aware Ride Assignment Algorithm} 
\label{sec:era}
\ERA is an offline algorithm (Algorithm~\ref{alg:alg_pseu}) designed to assign a set of $N$ riders to a fleet of $M$ drivers with the primary objective of minimizing emissions. The algorithm initialization phase (Lines 1-3) starts with an empty assignment for all $M$ drivers. Subsequently, the algorithm sequentially considers each passenger $n$ and generates all possible assignments of that passenger to the $M$ available drivers (Line 7). 
Importantly, the new assignments do not modify assignments to previous passengers.

\begin{figure}[!t] 
      \removelatexerror
      \begin{algorithm}[H]
      \label{alg:alg_pseu}
        \SetCustomAlgoRuledWidth{0.45\textwidth}  
        \caption{\small ERA($N$, $M$)}
           \small
           Initialize $Q_i = \{\}$, \ for $i \in [1,2, ..., M]$\;
           initial\_assignment = $\{ Q_i |\ \forall i \in [1, 2,..., M] \}$\;
           $\mathcal{A} = \{\text{initial\_assignment}  \}$\;
           \For{$n \in [1, 2, ..., N]$}{
                $\mathcal{A}' = \{  \}$\;
           \For{$a \in \mathcal{A}$}{
                $\text{children} = \text{Children}(a, n) $\;
                $\mathcal{A}' \leftarrow \mathcal{A}' \cup \text{children}$\;
            }      
            $\text{emission}_{\min} =$ Minimum emission of assignments in $\mathcal{A}'$\;
            $\mathcal{A} \leftarrow \{a |a \in \mathcal{A}',  \text{emission}_h(a) \leq \text{emission}_{\min}\}$
        }
      \end{algorithm}
\end{figure}
\setlength{\textfloatsep}{0pt}
\setlength{\intextsep}{0pt}

The algorithm relies on a heuristic function denoted as $\text{emission}_h(a)$ to estimate the emission associated with a specific assignment and calculates the minimum emissions in assigning all passengers to the drivers. 
Specifically, for any assignment $n \to m$ in the partial assignment $a$, the passenger $n$ is obligatorily assigned to driver $m$. 
We conduct an extensive numerical analysis \ERA to define a heuristic function, $\text{emission}_h(a)$. It {\em estimates} the emissions generated by the system based on the assignments made in a partial assignment $a$. The function estimates the emissions by calculating the deadhead distance for each driver. The deadhead distance is the distance between the drop-off location of a passenger and the pick-up location of the next passenger assigned to the same driver. In addition, the algorithm takes into account the lower bound for the deadhead distance of any unassigned passenger $n$. This lower bound is calculated by finding the minimum distance required to drive to the pick-up location of passenger $n$ from the drop-off location of any passenger that arrived before $n$ in the assignment sequence. By deriving the lower bound on the minimum required deadhead distance for each unassigned passenger and leveraging the known emission rate per mile for each driver, the algorithm can effectively estimate the minimum emission for each ride. The total estimated emission is calculated as the sum of the emissions associated with the assigned passengers and the minimum emissions of the unassigned passengers, all for the partial assignment $a$.

\begin{figure}
    \centering
    \includegraphics[width=0.8\linewidth]{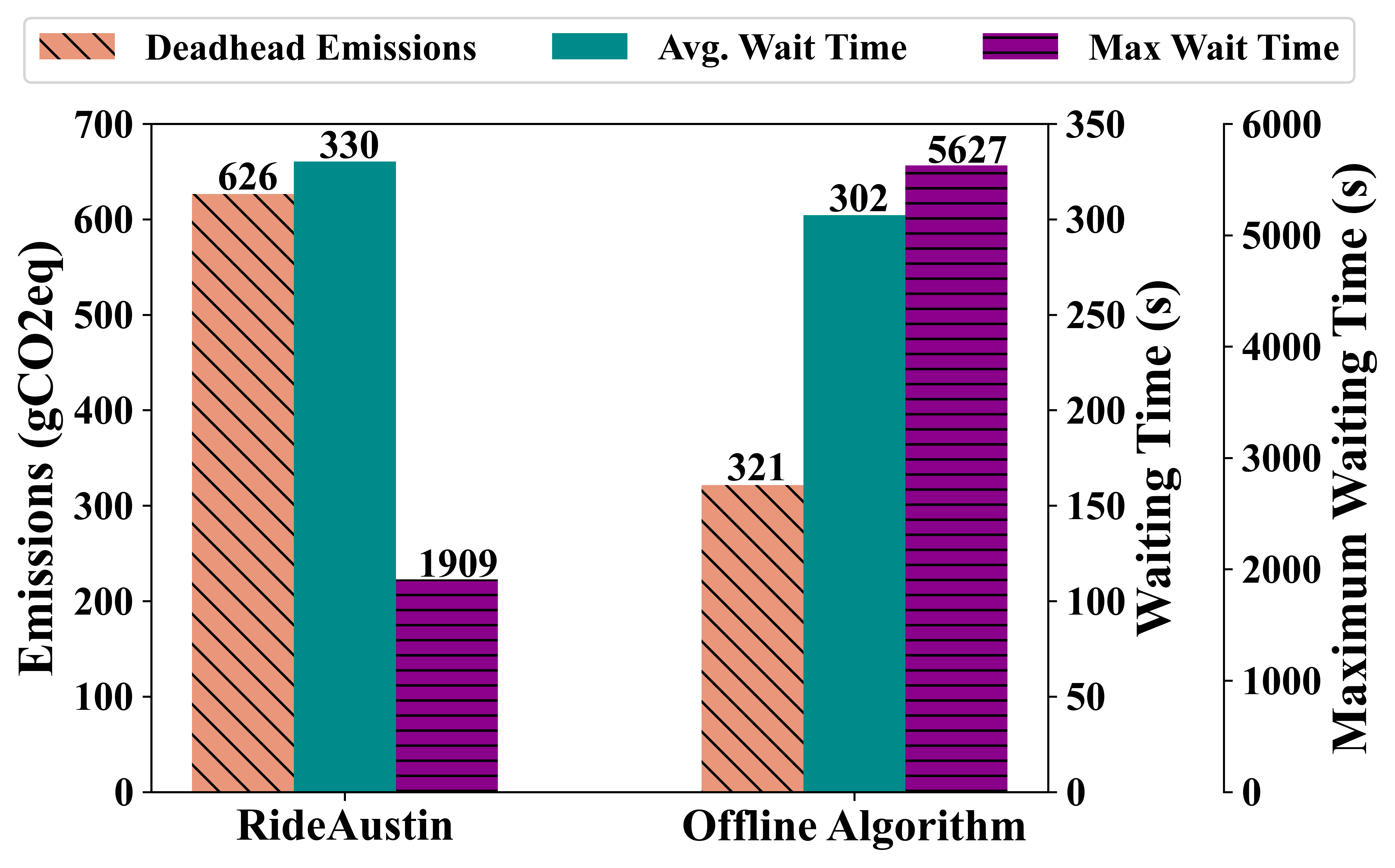}
    \vspace{-8pt}
    \caption{\emph{Opportunity analysis: Comparison of emissions from deadhead miles and waiting time for the default ride assignment and our offline emission‑aware ride assignment.}}
    \label{fig:pre_result}
\end{figure}

Finally, the algorithm keeps the assignments with the lowest estimated emission and starts branching from those (Lines 10-11). 
In this algorithm, $\text{Children}(\text{assignment}_{n-1}, n)$ returns a set of possible assignments of passenger index $n$ to drivers without changing any previously assigned passengers, $\text{assignment}_{n-1}$, (Line~7).
In Figure~\ref{fig:pre_result}, we present emissions from the deadhead miles and passenger waiting 
times for the current assignments in the dataset and the new assignments based on our emission-aware offline algorithm. 

We note that the computational complexity of the \ERA algorithm is substantial, rendering it impractical to seek near-optimal solutions for datasets with hundred thousands of trips. Consequently, for this preliminary feasibility study, we have limited our experimentation to a curated dataset comprising 142 samples. We randomly sampled 142 trips, completed by 14 unique drivers, from the RideAustin dataset~\cite{rideaustin-dataset} (detailed on Section~\ref{sec:exp_dataset}) on December 2, 2016. 
The left $y$-axis presents the total carbon emissions, measured in grams of carbon dioxide equivalent (g.CO2eq), for the deadhead miles of all the trips \footnote{This measurement is derived from multiplying the miles traveled by the fuel efficiency of the vehicle, denoted in grams of CO2 emitted per kilometer (gCO2/km). See the implementation section for additional details.}.  
The $y$-axes on the right show the waiting time (seconds), measured as the time period between a passenger posting the request and the driver picking up at different scales. 
Our results demonstrate that our proposed \ERA algorithm, albeit offline with complete knowledge of future rides, can reduce the deadhead miles emissions by 48.7\% (from 626gCO2eq to 321gCO2eq). 
Importantly, the average waiting across all trips also decreased by 8.5\% (from 330s to 302s) as \ERA reduced the deadhead miles. Still, the longest waiting time increased by 2.94$\times$ to almost 94 minutes from the 32 minutes observed for the default ride assignment. 

The above initial results show the potential of carbon emissions reduction by changing the assignment objective. However, \ERA is an offline algorithm; hence, it is not practically a relevant choice. Hence, in what follows, we present \TORA as an online and practical algorithm for emission-aware ride assignment. 

\subsection{\TORA: An Online Threshold-based Ride Assignment Algorithm}
\label{sec:tora}
\TORA is an online ride-assignment algorithm that controls the trade-off between waiting time and vehicle carbon emissions. To minimize waiting time for the passenger $n$, \TORA first finds the closest available driver to the passenger and then compares the distance and deadhead emission produced by other available drivers with the closest driver. \TORA calculates the \textit{Emission-to-Distance} (\texttt{E2D}) ratio for every driver and selects the appropriate driver for the passenger based on \texttt{E2D} values. More specifically, \texttt{E2D} is defined as the ratio between the difference of the deadhead emission of a driver and the deadhead emission of the closest driver over the difference of the distance of the passenger to the two drivers, i.e., 
\begin{equation}
    \label{eq:e2d}
    \texttt{E2D}(m, c) := \frac{e_d(n, c) - e_d(n, m)}{\text{distance}({m}) - \text{distance}({c})},
\end{equation}
where $\text{distance}(c)$, and $\text{distance}(m)$ show the distance of passenger $n$ to the closest driver $c$ and driver $m$.

The pseudocode for \TORA is outlined in Algorithm~\ref{alg:assign_vehicle}. It leverages a parameter, $\phi$, to balance the trade-off between reducing passenger waiting time and minimizing vehicle deadhead emissions. Initially, \TORA identifies the closest available driver and computes \texttt{E2D} value for other available drivers (Lines 1-2). Next, it picks a driver with the highest \texttt{E2D} ratio such that \texttt{E2D} $>$ $\phi \times E_0$, where $E_0$ are the emissions for a baseline vehicle over a unit distance trip. If such a driver exists, \TORA selects and returns it (Lines 6-7). Otherwise, it defaults to assigning the closest available driver.  

\begin{figure}[!t]
    \removelatexerror
    \begin{algorithm}[H]
        \label{alg:assign_vehicle}
        \SetCustomAlgoRuledWidth{0.45\textwidth}
        \caption{\small TORA(n, $\phi$)}
        \small   
        $c \gets$ closest available driver to n\;

        $m \gets \arg\max_{m'} \texttt{E2D}(m', c)$\;
        
        \If{ $\phi E_0 < \texttt{E2D}(m, c)$}{
            return $m$\;
        }
        \Else{
            return $c$\;
        }
    \end{algorithm}
\end{figure}
\setlength{\textfloatsep}{2pt}
\setlength{\intextsep}{0pt}

\begin{figure*}[t]
    \centering
    \begin{tabular}{ccc}
    \includegraphics[width=0.3\linewidth]{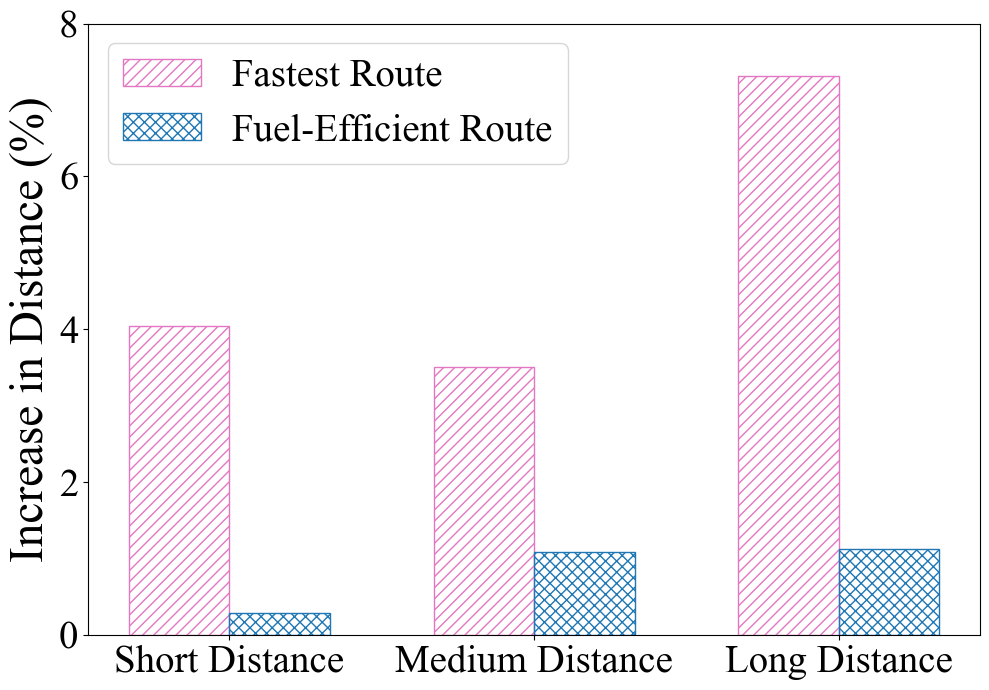} &
    \includegraphics[width=0.3\linewidth]{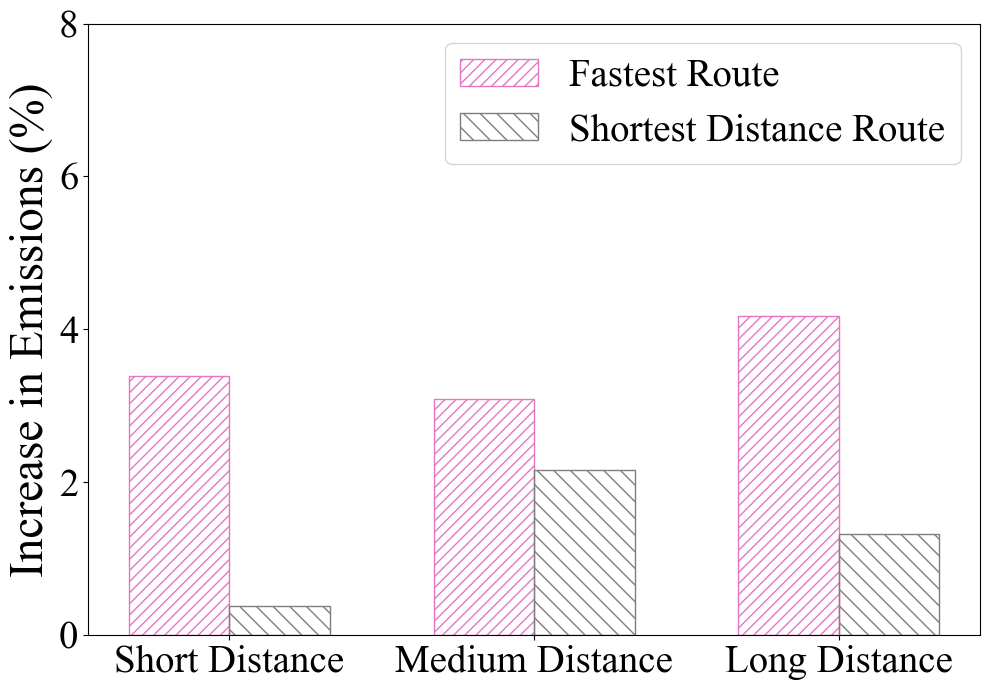} & \includegraphics[width=0.3\linewidth]{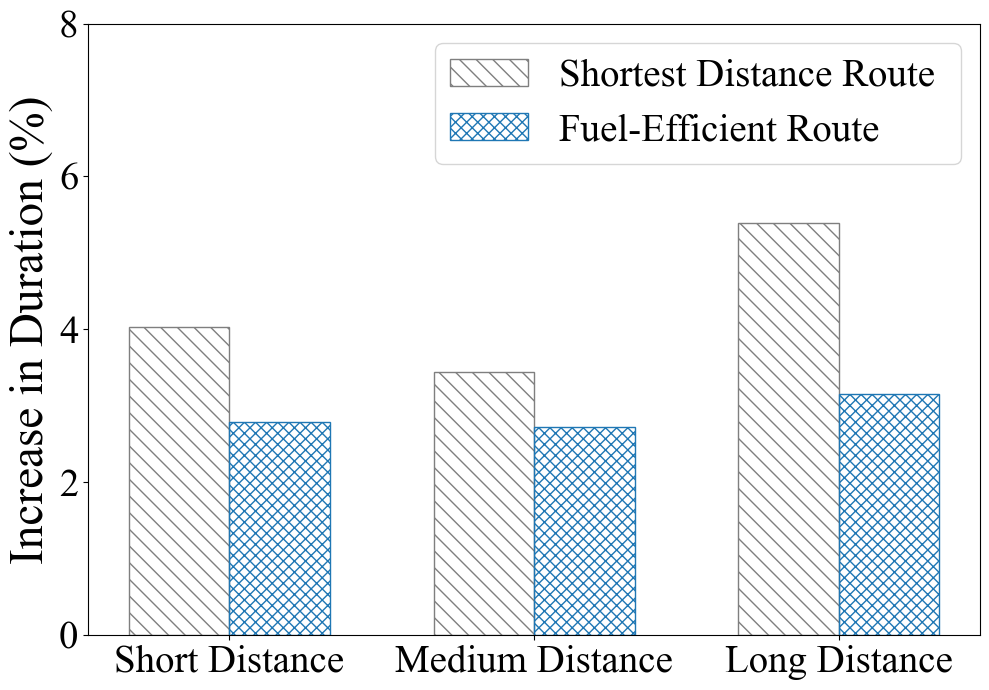}\\
    (a) Impact on Trip Distance  &  (b) Impact on Trip Emissions  & (c) Impact on Trip Duration
    \end{tabular}
    \vspace{-0.35cm}
    \caption{\emph{Average increase in distance (a), emission (b), and duration (c) for different routing algorithms to the optimal routes for short, medium, and long-distance trips for three trip categories in our analysis. }}
    \vspace{-0.3cm}
    \label{fig:routing_analysis}
\end{figure*}

The underlying concept of this algorithm is straightforward: it replaces the nearest driver with a more distant one if the latter emits $\phi E_0$ fewer emissions per additional unit distance traveled. The lower values of $\phi$ prioritize drivers with lower emissions, whereas higher values prioritize selecting the closest driver, potentially increasing emissions. To better understand how different values of $\phi$ impact the passenger's waiting time, we can calculate the upper bound on the distance of driver $m$ to the passenger when \TORA is willing to replace the closest driver $c$ by $m$. From the definition of \texttt{E2D}$(m, c)$ in Equation~\ref{eq:e2d}, \TORA replaces closest driver $c$ by driver $m$ if
\begin{equation}
\label{eq:phi}
    \phi E_0 < \frac{e_c\cdot \text{distance}(c) - e_m \cdot\text{distance}(m)}{\text{distance}(m) - \text{distance}(c)},
\end{equation}
where $e_c$ and $e_m$ denote the unit-distance emission for the driver $c$ and $m$. By rearranging Eq.~\eqref{eq:phi}, \TORA replace driver $c$ with driver $m$ if
\begin{equation}
    \label{eq:TORA_dist_lim}
    \frac{\text{distance}(m)}{\text{distance}(c)} < \frac{e_c / E_o + \phi}{e_m/E_0 + \phi},
\end{equation}
where the ratio of $\text{distance}(m) /\text{distance}(c)$ approximates the ratio of waiting time for the passenger with driver $m$ over driver $c$. Eq.~\eqref{eq:TORA_dist_lim} shows that \TORA limits the distance of the lower emission driver based on its and closest driver's unit-distance emissions.


%% file: 4-routing.tex
In this section, we study the impact of different route options on the total distance, trip duration, and emission produced during the trip. We consider the shortest route, fastest route, and most fuel-efficient route as the three possible choices of driver to travel from pick-up to drop-off location of passenger. We collected the data for the three route options --- distance, duration, and fuel consumption --- for 44,794 trips using the Google Maps API~\cite{gmapapi}. In our analysis, we categorize trips into short- (<1 mile), medium-(1--10 miles), and long-distance (>10 miles) trips (based on the length of shortest route) to emphasize the impact of different routing options on different trips. Our three route options also give us three metrics for our analysis: distance traveled, time taken, and emissions produced. We choose one route option as a baseline (optimize one metric) and evaluate how worse off the other two routes are based on the baseline metric for all trip categories.




In Figure~\ref{fig:routing_analysis}(a), we examine the percentage increase in distance for the fastest and fuel-efficient route options when compared to the baseline shortest distance route ($y$-axis) for the three trip categories ($x$-axis). The fastest route takes a longer route, probably to avoid traffic congestion, by up to 7.5\%, depending on the length of the trip. The fuel-efficient route needs to balance the savings from traveling the shortest distance and the cost of idling in a traffic jam; it drives longer highly judiciously. Figure~\ref{fig:routing_analysis}(b) shows the increase in trip emissions for the shortest distance and fastest routes. The increase in emissions is smaller than the increase in distance we observed in Fig.~\ref{fig:routing_analysis}(b). Also, the actual magnitude of increase even for the fastest trip is around 4\%. The differences may further diminish in practice, especially in cities where traffic patterns change quickly and unexpectedly. Finally, in Figure~\ref{fig:routing_analysis}(c) we evaluate the time cost of the shortest distance and fuel-efficient route options. As we observed before, shortest distance routes are most likely congested, and choosing them would cost up to 6\% more time than the shortest duration route. Since the fuel-efficient route option travels longer distances but avoids congestion, its increase in time is smaller. 

Our analysis showed that the choice of route had varying impacts on trip duration, distance, and emissions, with the shortest, fastest, and fuel-efficient routes exhibiting different trade-offs.

\noindent
\emph{\textbf{Key takeaways.} 
Our analysis yields a few interesting and surprising insights. It shows that choosing the fuel-efficient route may not always yield huge fuel savings. However, since the distance and time increase for the fuel-efficient trip are minute, $<$1\% and $<$2.5\%, respectively, the small savings do not come at a huge cost. Also, the results are subject to local road networks, traffic conditions, and driver habits.  
}

%% file: 5-toolkit_description.tex

Our work on developing emission-aware ride assignment and routing algorithms requires ridesharing datasets to include carbon emission information for the deadhead miles and individual trips, and equity information on the drivers and riders.
However, no existing ridesharing datasets contain such information. Augmenting the existing datasets with emission and equity information requires solving multiple challenges. 
In this section, we present the design and implementation \kit\ \footnote{\href{https://github.com/Mahsahebdel/e2\_ridekit}{https://github.com/Mahsahebdel/e2\_ridekit}} for augmenting ridesharing datasets as an independent contribution. To make the toolkit complete and independently useful, we implement several additional functionalities not used directly for our algorithmic work in this paper. However, they are of independent interest for emissions research in the ridesharing ecosystems. 


\subsection{\kit\ Design}
\label{sec:design}

The design of \kit\ needs to solve three key challenges: (1) building vehicle- and congestion-aware emission models for internal combustion engine (ICE) powered vehicles, (2) modeling congestion-aware discharging and location-aware charging models for Electric Vehicles (EVs), and (3) defining, quantifying, and adding missing information on equity-related metrics. 
We solve these challenges as a part of our design, which we describe next. 

\begin{figure}[t]
    \centering
    \includegraphics[width=\linewidth]{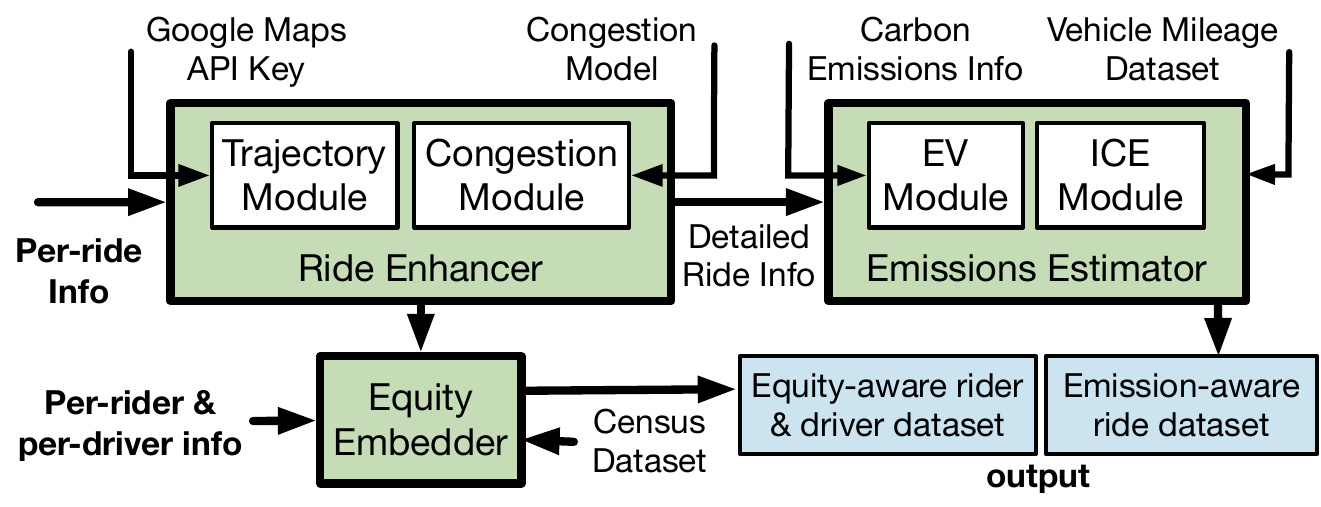}
    \caption{\emph{The high-level architecture, various components, and workflow of \kit.}}
    \label{fig:kit-arch}
    \vspace{10pt}
\end{figure}

\vspace{0.1cm}
\noindent
\textbf{1. ARCHITECTURE.} 
The high-level architecture of \kit\ consists of three major modules, as shown in Figure~\ref{fig:kit-arch}.
The three major modules are (i) \texttt{Ride Enhancer} module, (ii) \texttt{Emission Estimator} module, and (iii) \texttt{Equity Embedder} module.

\vspace{0.05cm}
\noindent
\textbf{\texttt{Ride Enhancer}.}
The goal of this module is to augment the per-ride/trip information from an existing dataset with missing information, if any. 
For example, a dataset may only provide the pickup and drop-off coordinates without sharing information on the actual route taken by the driver. Since we need to estimate per-ride emissions, which depend on the route the driver takes, \texttt{Ride Enhancer} module can leverage a navigation API to generate trajectories. 
Similarly, the congestion model inside the \texttt{Ride Enhancer} module is responsible for estimating congestion during the trip since it is one of the key factors that affect the fuel efficiency of vehicles, which impacts the carbon emissions for a given trip.
Note that \texttt{Ride Enhancer} module operates on all rides and does not differentiate between rides from an ICE-powered vehicle and an EV.

\vspace{0.05cm}
\noindent
\textbf{\texttt{Emission Estimator}.}
This module is the heart of the \kit\ as it estimates the carbon and other GHG emissions for various aspects of each ride. For example, it will estimate the emissions during the actual trip with the passenger and emissions from deadhead miles that the driver had to undertake to pick up a new customer. 
Since the car make/models, fuel types, and efficiencies of ICE-powered vehicles and EVs are significantly different, we use separate modules for estimating their emissions. 
Based on the vehicle type, EV or ICE-powered, we use data from either carbon information services such as electricityMaps~\cite{electricity-map} and Watttime~\cite{watttime} or existing vehicle emissions datasets~\cite{canadacar_emissions}.

\vspace{0.05cm}
\noindent
\textbf{\texttt{Equity Embedder}.}
An equally important and the most challenging module is the \texttt{Equity Embedder} module that embeds socioeconomic and demographic information for the drivers and the riders. 
Due to privacy concerns, dataset collectors and maintainers do not release information that can reveal the identity of the riders or passengers. 
As a result, this information needs to be estimated at a coarse granularity using publicly available information using resources such as US Census Bureau~\cite{us-census}. 
Specifically, we use the Geocode API \cite{geocodIO} to collect demographic data such as race, population, and median income. 
The data is provided by the U.S. Census Bureau and is made available by census block, i.e., the smallest geographical unit for which the U.S. Census Bureau collects and provides statistical data.
In some cases, the data may not be for the same year as the ride-sharing dataset.
For example, for many locations in the US, 2020 is the most recent year with available census data~\cite{Census:2020}. 
However, this should not be a significant issue as the demographics of cities, or even blocks, only change over the course of many years. 
Using per-ride information, our module optionally estimates coarse information on which segment of the city the driver or rider lives. 
This information can be gleaned from past datasets for the regular drivers and riders using methods such as K-means clustering~\cite{lloyd1982least}.
We release this module with a disclaimer about the potential inaccuracies in the embedded information.

\vspace{0.1cm}
\noindent
\textbf{2. WORKFLOW.}
We next describe the workflow of \kit\ when using it to augment an existing ridesharing dataset. 
We first split the input dataset into two segments that contain information on individual rides and the other segment containing information on all the drivers and riders if available. 

If the dataset does not provide trajectory information, we leverage navigation APIs with generative modeling to estimate trajectories for each ride, i.e., the route between pickup location and destination. 
This trajectory information is needed to estimate the trip emissions based on the vehicle used, route taken, and distance traveled. 
We next adjust the trip emissions based on the estimated travel time index (TTI), which compares the travel time of a given trip with the typical time for the same trip and represents a good proxy for traffic congestion. By leveraging the appropriate emission module, this step handles both ICE-powered vehicles and EVs.

To solve the second challenge, when considering discharging, we will leverage the same approach as the first challenge. 
To model the charging patterns and constraints, we will leverage publicly-available information on the charging station locations~\cite{charging-locations}, the ratings of the available chargers~\cite{charging-locations}, and the carbon intensity of electricity supply using services such as electricityMaps~\cite{electricity-map} and Watttime~\cite{watttime}.
To solve the final challenge, we will create a meta-dataset that provides the distributions of equity-related metrics augmented with additional information on socio-economic factors from U.S. Census Bureau~\cite{us-census}. Table~\ref{table:fields} lists the additional per-trip and meta fields that \kit add to our example dataset.

\subsection{\kit\ Implementation}

In this section, we describe the implementation of \kit\ and the models and datasets we use for our case study, as explained in the following.
We combine multiple additional datasets as a part of our implementation and describe them next.


\vspace{0.05cm}
\noindent\textit{1. Car emissions dataset.}
To model emissions from the ICE-powered vehicles in the dataset, we use the Canada Car Emissions dataset~\cite{canadacar_emissions}. This extensive and reliable resource provides a comprehensive overview of CO2 emissions data for various vehicles. 
This dataset covers a period of 7 years and is sourced from the official open data website of the Canadian government. 
It consists of detailed information on different aspects and characteristics of 26,075 unique vehicles. The dataset primarily focuses on essential factors such as vehicle's make, model, year, and carbon emissions, which are fundamental in understanding CO2 emissions. However, it goes beyond these key attributes and includes valuable information such as fuel type, engine size, transmission type, and other relevant features. 
Easy access to these additional details through the toolkit will enable researchers and analysts to gain deeper insights into the relationship between car attributes and carbon emissions, facilitating a more comprehensive dataset analysis.



\vspace{0.05cm}
\noindent\textit{2. Carbon emissions dataset.}
ElectricityMaps~\cite{electricity-map} and Watttime~\cite{watttime} are valuable services that provide essential information regarding the carbon intensity of electricity supply across different locations. 
This information is presented as average and marginal hourly values, which can vary depending on the specific location and time. The availability of this data is particularly beneficial for modeling the carbon intensity of electricity supply during the charging of EVs. 
To accurately model the charging patterns and constraints for EVs, publicly available data on charging station locations and charger ratings are utilized. 
This data aids in estimating the charging patterns of EVs and determining the time required for a vehicle to be fully charged. 
However, it is not always feasible to determine the exact time at which an EV is charged. 
In such cases, a proxy approach is employed, where the daily average carbon intensity value of the electricity supply at the location where the vehicle is charged is utilized to estimate the charging time. 
By adopting this approach, an estimation of the carbon intensity of the electricity supply at the time of charging is derived and incorporated into the analysis. 
By leveraging the information from electricityMaps, WattTime, and the data on charging station locations and charger ratings, researchers and analysts can enhance their analysis of the environmental impact of EVs and their charging patterns. 
This integration allows for a holistic understanding of the relationship between EV charging, the carbon intensity of electricity, and its implications for sustainability and environmental considerations.

\vspace{0.05cm}
\noindent\textit{3. Traffic congestion model.}
The traffic congestion model is a demand modification and traffic simulation process.  The model aims to estimate changes in travel demand and traffic patterns. We next detail the demand modification and traffic simulation processes.

In the demand modification process, we adjust the Origin-Destination (O-D) matrix to reflect changes in travel demand resulting from various factors. This requires data-derived inputs, such as trip proportions associated with specific Traffic Analysis Zones (TAZ) and the proportion of dedicated tours for specific purposes. Users can also set additional simulation parameters, like the penetration and substitution rates. To incorporate changes in travel patterns, we use a Monte Carlo simulation, which accounts for uncertainties in determining trip categories and subsequent destinations by generating multiple potential scenarios. After identifying the affected trips, the OD matrix is modified to reflect the changes introduced in the simulation. 

The traffic simulation process involves using the Vehicle Hours Traveled (VHT) metric to understand and quantify the impact of transportation activities on congestion. VHT measures the total travel time of vehicles on the road network during a specific period, considering both the number of vehicles and the duration of their trips. To calculate VHT, we use Static Traffic Assignment, which assigns trips from an Origin-Destination (OD) matrix to the transportation network, considering travel demand, road capacities, and traffic conditions. The goal is to distribute the trips across the network in a way that reflects realistic travel patterns and estimates travel times for the assigned trips. 
The resulting network data provides valuable information on updated link speeds, travel times, and vehicle volumes, enabling the calculation of VHT by summing up the travel times on each link. By comparing VHT values and analyzing different scenarios, researchers can assess the impact of transportation activities on congestion, aiding in the understanding of traffic dynamics, evaluation of interventions, and development of strategies to mitigate congestion-related issues.

\begin{table}[t]
\scriptsize
    \caption{Additional Fields Added to the RideAustin Dataset.}
    \vspace{-0.2cm}
    \begin{center}
    \begin{tabular}{| l | c | c | c |} \hline
    \textbf{Trip related fields} &  fuel efficiency & emissions (deadhead \& trip) & congestion index \\ \hline   
    & EV (yes/no) & demographics (rider \& driver) & \\ \hline  
    \textbf{Meta fields}  & EV trajectories & electricity carbon intensity & charger location \\ \hline 
    & charger type & charging/discharging model & \\ \hline 
    \end{tabular}
    \label{table:fields}
    \end{center}
    \vspace{-0.0cm}
\end{table}

%% file: 6-experiments.tex
This section provides a comprehensive experimental study to evaluate the deadhead, total emission, and waiting time of a ridesharing platform under ride assignment of \TORA with different $\phi$ values and different routing algorithms explained in Section~\ref{sec:routing_tradeOff}. The key questions for the evaluations and our findings are outlined below. 

\begin{enumerate}[leftmargin=*, topsep=-0.02cm]
    \item[Q1] \textit{How does \TORA impact the deadhead emissions and waiting times when the routing algorithm remains unchanged?} 
    We find that \TORA can improve the deadhead emission significantly with a cost of slightly increasing the waiting time (Key takeaway 1).
    \item[Q2] \textit{How does \TORA affect the equity among drivers with varying vehicle emissions?} Our analysis reveals that reduction in emissions requires sacrificing equity by assigning more rides to low-emission vehicles (Key takeaway 2).
    \item[Q3] \textit{How do the \TORA and routing algorithms compare regarding their impact on the emission of ridesharing systems?} We find that the impact of the ride assignment algorithm on the emission and waiting time of the ridesharing system is substantially more than the impact of routing algorithms (Key takeaway 3).
\end{enumerate}




  

\subsection{Experimental Setup}
\label{sec:exp_setup}

We outline the characteristics of the ridesharing dataset we use, various parameters that we vary in our experiments, and the metrics we use to evaluate the efficacy of our approach in reducing emissions and ensuring equity.



\noindent
\textbf{Ridesharing dataset.}
\label{sec:exp_dataset}
RideAustin, a non-profit ridesharing company based in Austin, Texas, has released a dataset encompassing around 1.5 million trips spanning a 10-month period across 2016 and 2017~\cite{rideaustin-dataset}. During this period, no other major ridesharing platforms like Uber or Lyft operated in Austin. The dataset contains comprehensive trip-specific information, including the trip's start and end times, geographical coordinates for both pick-up and drop-off locations, the vehicle's make and model, distance covered before, during, and after each trip, and unique identification numbers for the driver and passenger.

In our experiments, we use RideAustin dataset consisting of trips between December 1, 2016, and December 10, 2016. This subset comprises 44,794 ridesharing trips conducted by 1,406 drivers. We augment this dataset with emissions information using \kit, and the additional dataset explained in Section~\ref{sec:toolkit}. We categorized the vehicles emitting less than 135 g.CO2eq/km as low-emission vehicles (LEVs), and created three more variants of the dataset by randomly selecting 5\%, 10\%, and 20\% of the vehicles which are not categorized as a low-emission and converting them into electric vehicles (EVs), yielding three addional datasets with 10\%, 15\%, and 25\% low-emission vehicles. During our experiments, we set the unit distance emission equivalent of EVs to be 63.35 g.CO2eq/km\footnote{We used the Tesla Model Y as a prototype EV, which has an energy efficiency rating of 26 kWh/100mi. We used the average carbon intensity value for Austin, Texas, which is 408 g.CO2eq/kWh to compute the unit distance emissions.}.

\noindent
\textbf{Baseline strategies and parameter ranges.} For our ride assignment algorithms, we use two baseline ride assignment strategies: 1) default assignment as presented in the RideAustin dataset and 2) shortest distance assignment that assigns the closest driver to the passenger. The second assignment strategy is emulated as an extreme case of \TORA when $\phi$ approaches infinity. We use three variants of \TORA with $\phi$ values of 0.1 (representing an emissions-prioritized assignment), 1 (roughly representing emissions- and -waiting-time-aware assignment), and 7.5 (representing a waiting-time-prioritized assignment) for our ride assignment strategies. For routing, we used three routing strategies: the shortest, the fastest, and the fuel-efficient, as explained in Section~\ref{sec:routing_tradeOff}. 

\begin{figure}[t]
  \centering
    \begin{tabular}{cc}
    \includegraphics[width=0.47\linewidth]{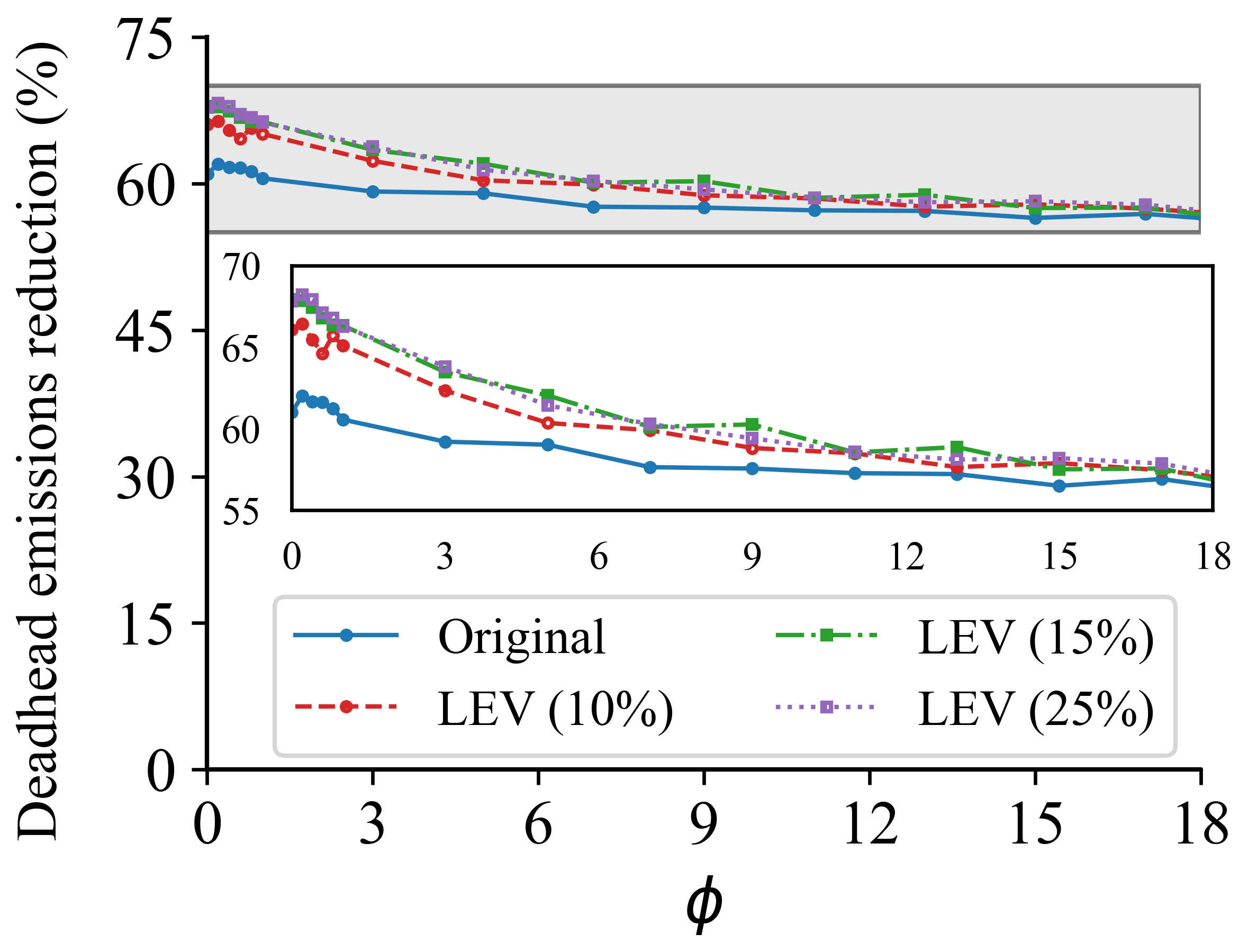} &
    \includegraphics[width=0.47\linewidth]{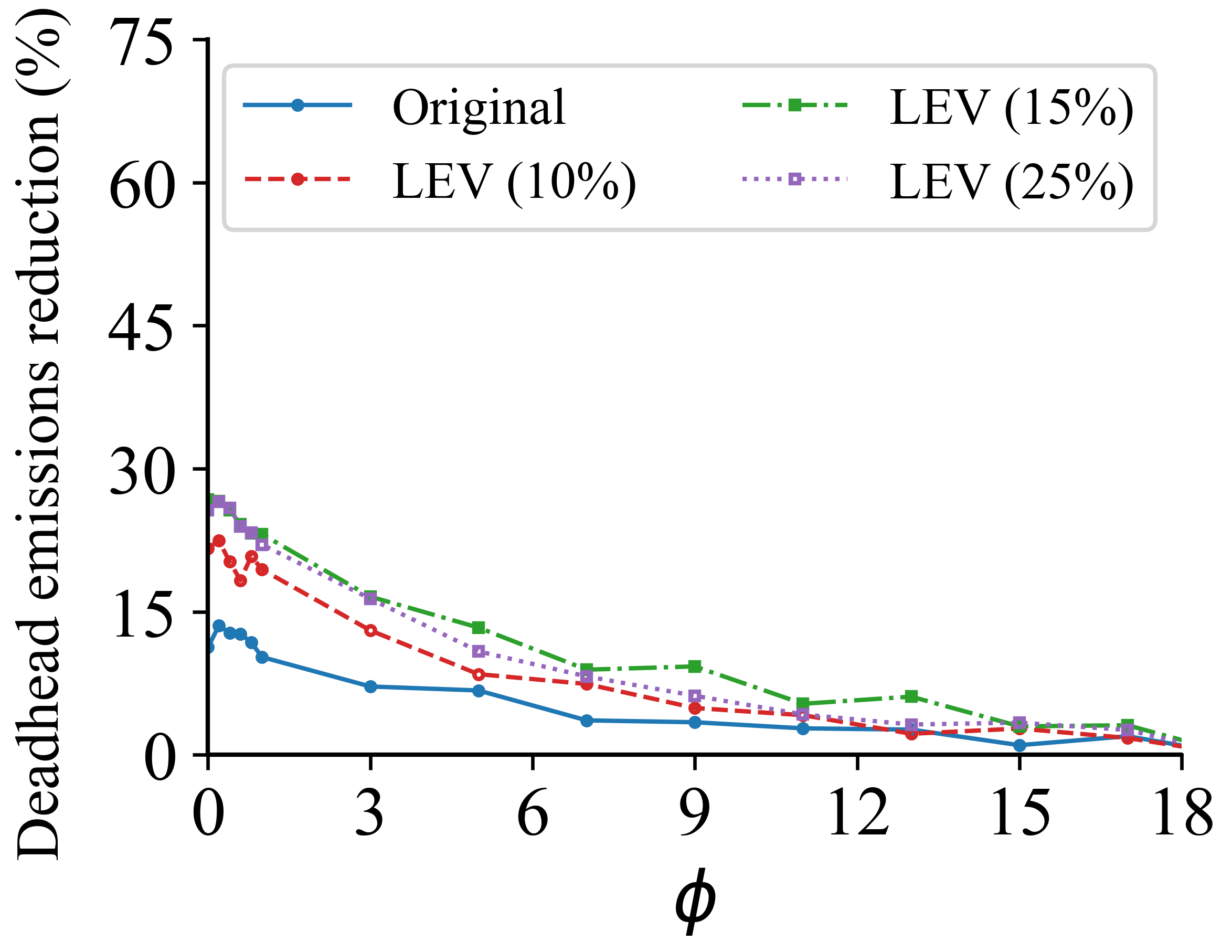} 
    \vspace{-0.08cm}
    \\
    (a) Default Assignment &  (b) Shortest Distance 
    \end{tabular}
  \caption{Percentage reduction in deadhead miles emissions on $y$-axis compared to the (a) default assignment in the RideAustin dataset and (b) shortest distance driver assignment, as a function of $\phi$ on the $x$-axis for different LEV fractions. Lower values of $\phi$ generally yield higher reduction.
  }
  \vspace{10pt}
  \label{fig:deadh_assignment}   
\end{figure}

For our experimental parameters,  we vary the values of $\phi$ between 0.001 and 18. We set the value of $E_0$ to 63.35 g.CO2eq/km, which is equal to the unit-distance carbon emission for an EV. It is worth noting that scaling up the value of $E_0$ is equivalent to scaling down the value of $\phi$. For example, the ride assignment of \TORA with $\phi = 10$ and $E_0 = 63.35 g.CO2eq/km$ equals to the ride assignment of \TORA with $\phi = 5$ and $E_0 = 126.7 g.CO2eq/km$.

\noindent
\textbf{Performance and equity metrics.}
We assess ride assignment strategies using three performance metrics: passenger waiting time, deadhead emissions, and overall ridesharing emissions. 

To analyze the equity implications of the ride assignments algorithms, we leverage the parameter $\phi$. Besides the previous LEVs, we introduce another category of high-emission vehicles (HEVs) with lower than 11.7 liters per 100km (20mpg).
We report the fraction of trips assigned and the average deadhead-to-trip distance ratio for the two categories. For an equitable outcome, the values for both metrics should be similar across vehicle categories.  

\subsection{Impact on Emissions and Waiting Times}

\begin{figure}[t]
  \centering
  \begin{tabular}{cc}
  \includegraphics[width=0.47\linewidth]{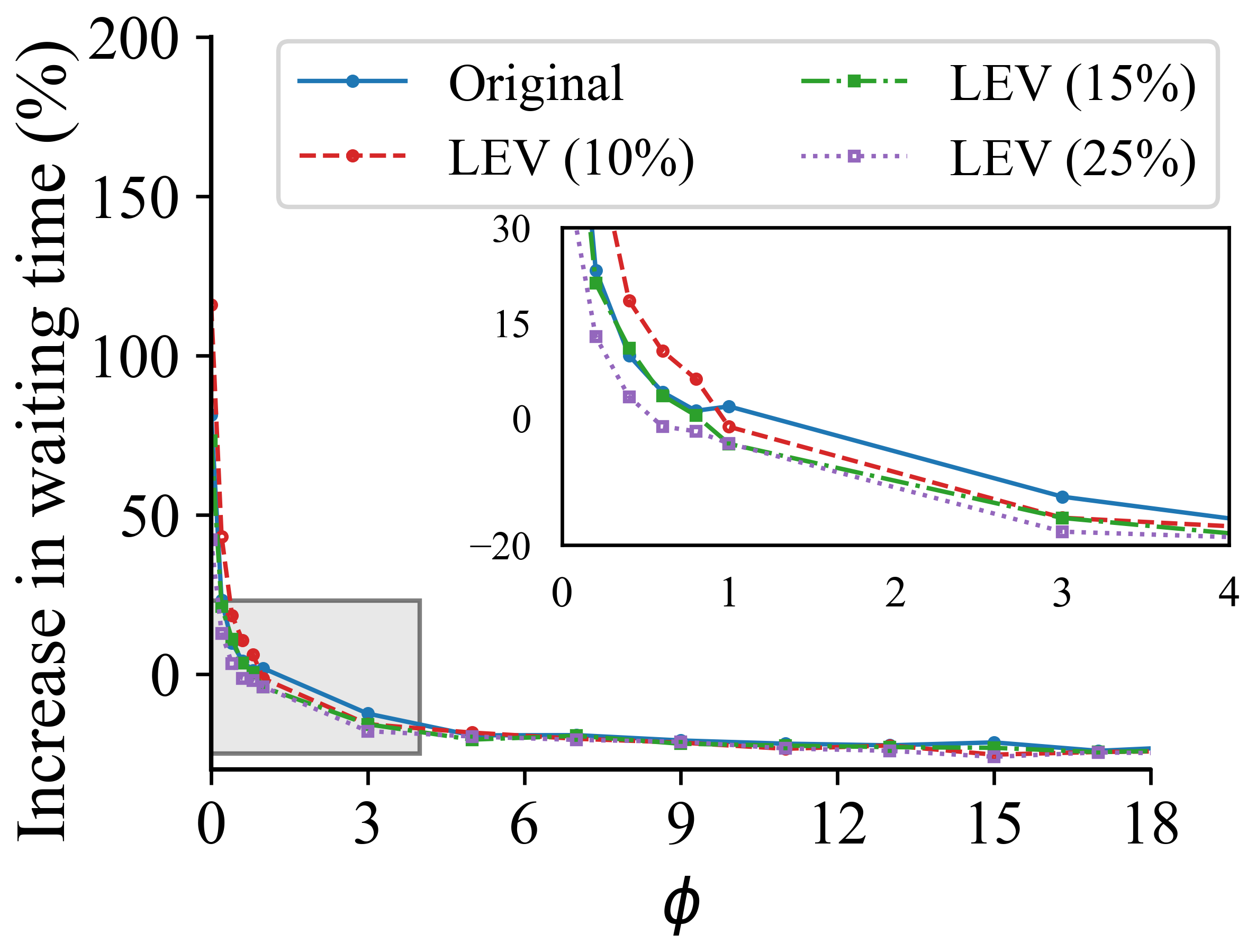} &
  \includegraphics[width=0.47\linewidth]{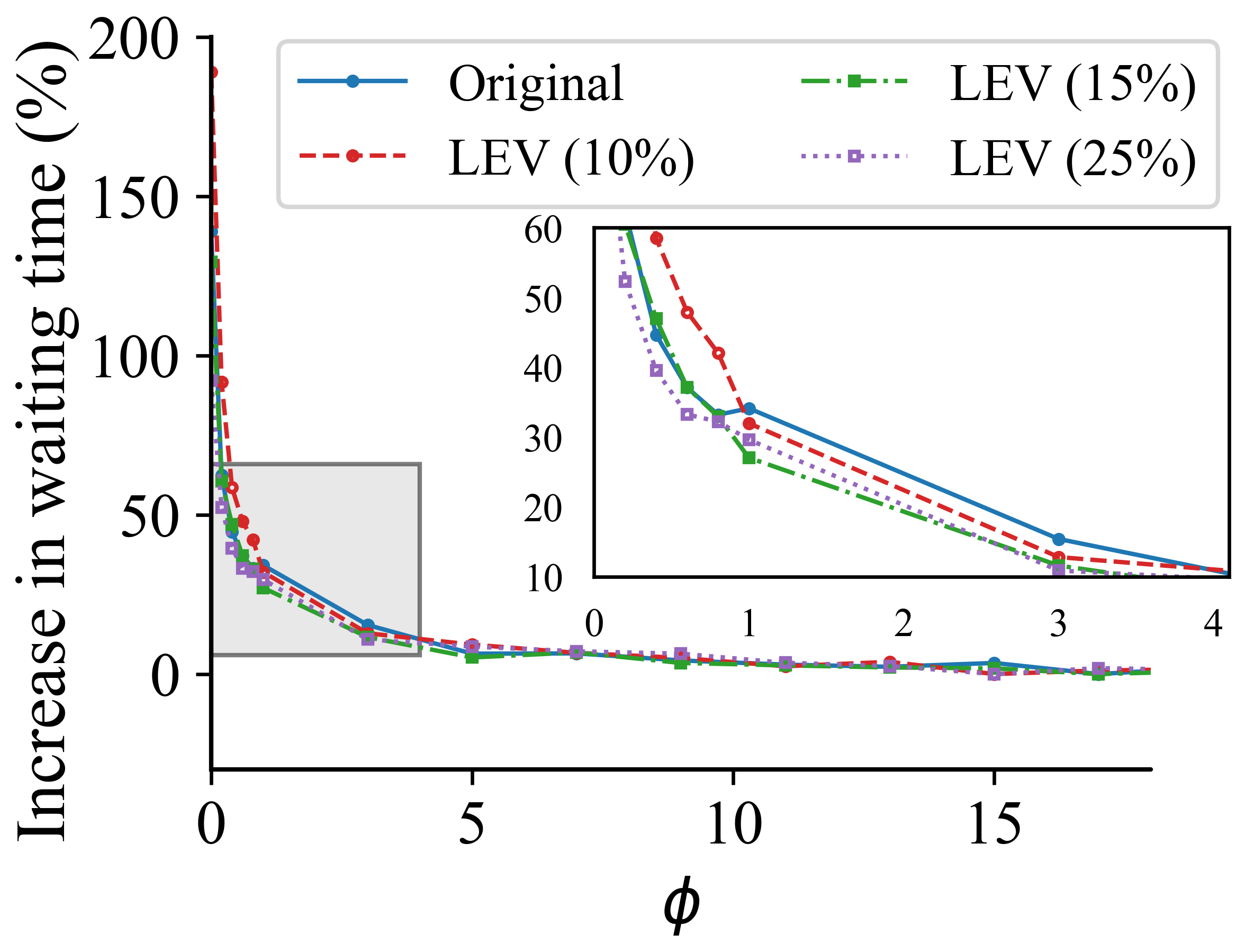} 
  \vspace{-0.08cm}
  \\
  (a) Default Assignment &  (b) Shortest Distance 
  \end{tabular}
  \caption{
  Percentage increase in waiting time on the $y$-axis compared to the (a) default assignment in the RideAustin dataset and (b) shortest distance driver assignment, as a function of $\phi$ on the $x$-axis for different LEV fractions. Higher values of $\phi$ generally yield lower waiting time.}
  \vspace{10pt}
  \label{fig:waiting_assignment}   
\end{figure}

In this section, we present the outcomes of our experiments, which shed light on the intricate interplay of ride assignment algorithms, routing strategies, and various dataset configurations. These results offer valuable insights into the performance trade-offs inherent in ridesharing platforms. We report our results by varying the threshold values ($\phi$) for different datasets, including the original vehicle dataset and synthetically generated datasets with the injection of 10\%, 15\%, and 25\% low-emission vehicles (LEVs). The objective is to investigate the impact of threshold parameter $\phi$ on two pivotal factors: the reductions in deadhead emissions (in Figure \ref{fig:deadh_assignment}) and the increase in waiting times of the riders (in Figure \ref{fig:waiting_assignment}). 

The results in Figure \ref{fig:deadh_assignment} show that smaller values of threshold value $\phi$ lead to greater improvements in deadhead emissions. Additionally, at a fixed threshold value, a higher percentage of low-emission vehicles results in comparatively greater improvements in deadhead emissions. Regarding waiting times, as shown in Figure~\ref{fig:waiting_assignment}, a smaller threshold value increases the relative waiting time for both the default assignment and the shortest distance assignment strategies. Also, as $\phi$ increases, the relative increase in waiting time decreases. We note that with higher values of $\phi$, our results demonstrate a negative increase in waiting time, i.e., a reduction in waiting time, compared to the default ride assignment in the RideAustin dataset. This happens because the default assignment does not always assign the closest drive to the passenger, and there is an opportunity to reduce waiting time without increasing emissions. 

\vspace{0.05cm}
\noindent 
\emph{\textbf{Key takeaway 1.} There is a trade-off between reducing the average deadhead emission and reducing the average passenger's waiting time under the ride assignment of \TORA, e.g., while smaller threshold values substantially reduce the emissions, it comes at the expense of increased waiting time of riders. }


\subsection{Equity Implications}
Depending on how we look at the problem and who we ask, the definition of equity can significantly change. In our context, we define it as an equitable distribution of rides across all vehicles. Since emission-aware ride assignment strategies favor high-efficiency vehicles, they can starve drivers with low efficiency, which typically belong to low-income drivers. As a result, there is a tradeoff between reduction in emissions and equity. In this section, we navigate the tradeoff using various values of $\phi$. Note that any other definition of equity, such as the waiting time experienced by riders from low-income communities vs. affluent neighborhoods, can be explored using the information augmented into the dataset by \kit. 

\begin{figure}[t]
  \centering
  \begin{tabular}{cc}
  \includegraphics[width=0.48\linewidth]{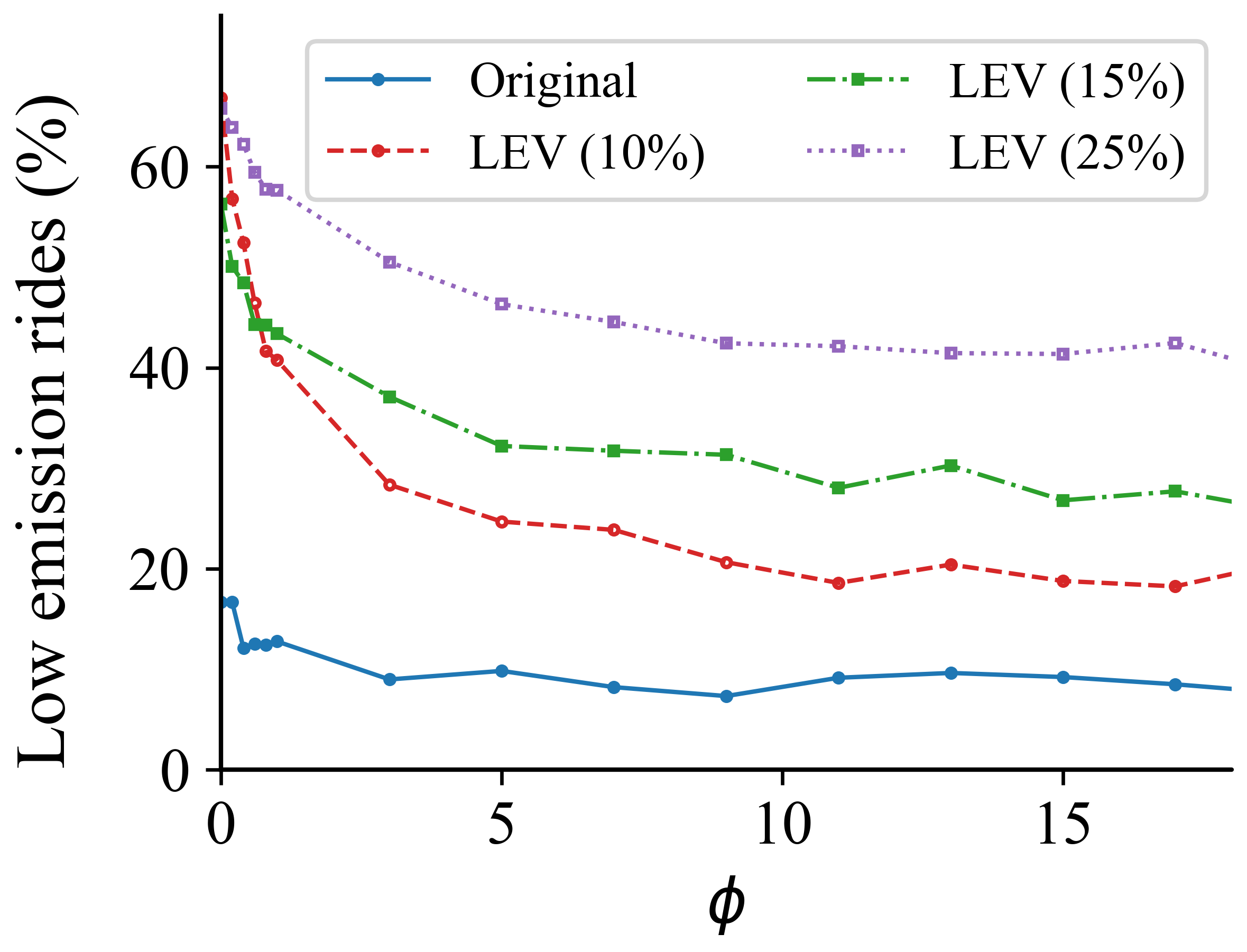} &
  \includegraphics[width=0.48\linewidth]{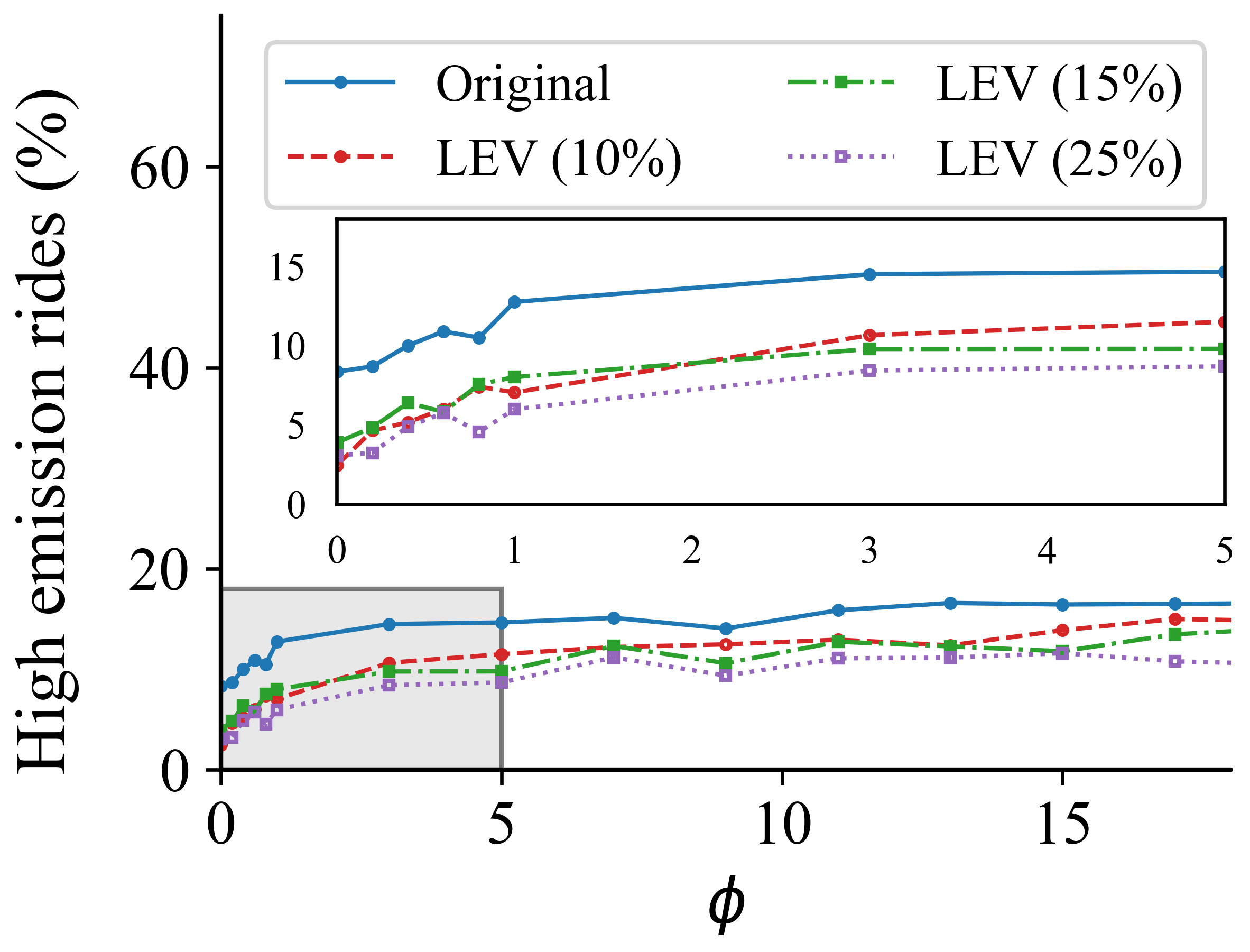} 
  \vspace{-0.08cm}
  \\
  (a) LEVs &  (b) HEVs
  \end{tabular}
  \vspace{-0.4cm}
  \caption{
  Fraction of rides assigned on $y$-axis to (a) LEVs and (b) HEVs, as a function of $\phi$ on the $x$-axis for different LEV fractions. At lower $\phi$ values, \TORA  assigns more rides to LEVs.}
  
  \label{fig:equity_fraction}   
\end{figure}

As illustrated in Figure~\ref{fig:equity_fraction}, adjusting the threshold value results in a trade-off: an increase in the threshold leads to a decrease in the percentage of trips assigned to LEVs, while the percentage of rides assigned to HEVs increases. This implies that a higher threshold favors HEV rides, a trend consistent across datasets with varying LEV percentages. Intriguingly, our results also demonstrate that a higher percentage of LEVs within a dataset corresponds to a greater percentage of trips completed by LEVs, irrespective of the threshold value. Notably, the original dataset, across a range of threshold values from $0.001$ to $18$, assigned between $7.6\%$ to $16.6\%$ of rides to LEVs. These percentages shifted to $20.7\%$ to $66.8\%$, $25.5\%$ to $56.2\%$, and $39.4\%$ to $65.8\%$ for datasets containing 10\%, 15\%, and 25\% LEVs, respectively. This underscores the pivotal role of a higher proportion of LEVs in achieving environmentally friendly ridesharing outcomes within defined criteria. 

\begin{figure}[t]
  \centering
  \begin{tabular}{cc}
  \includegraphics[width=0.48\linewidth]{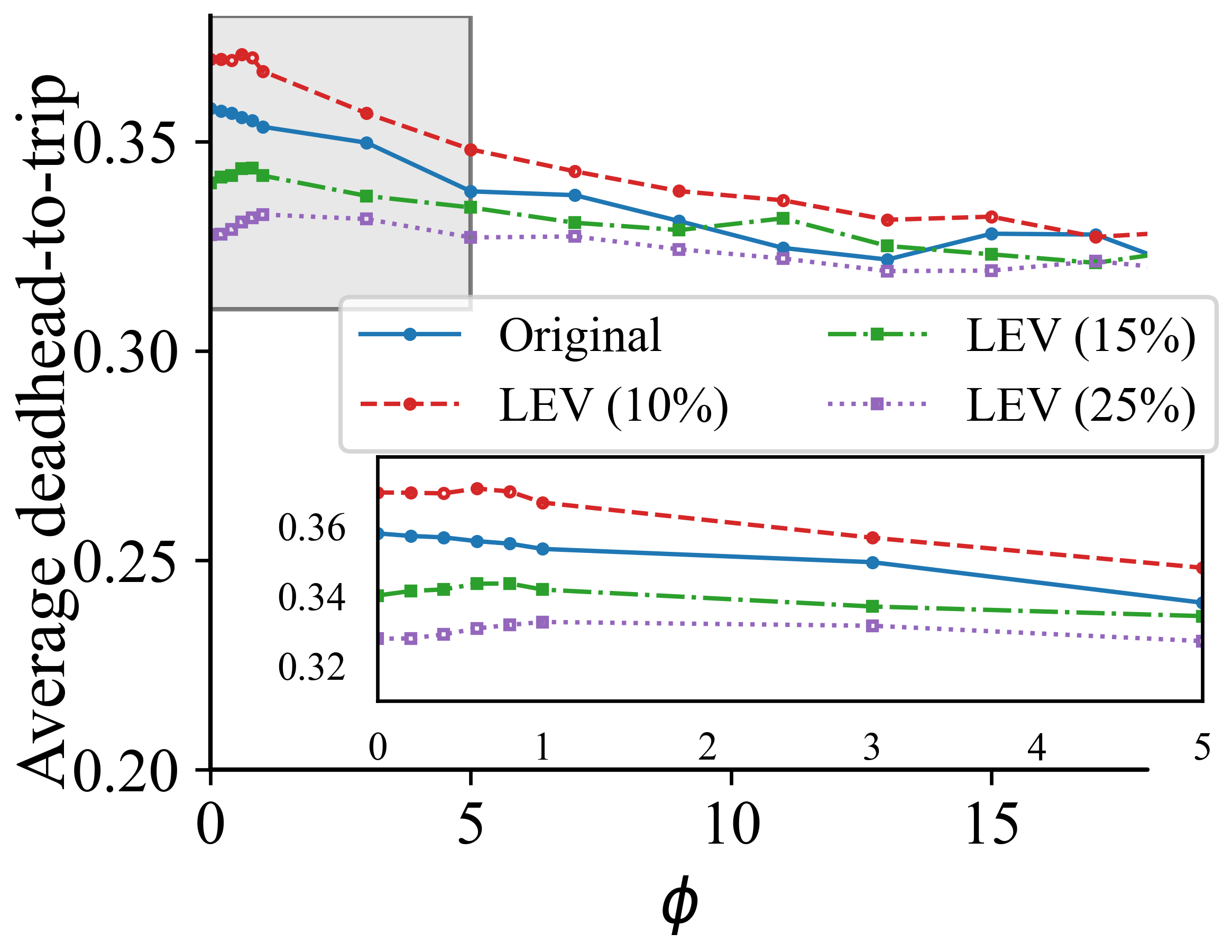} &
  \includegraphics[width=0.48\linewidth]{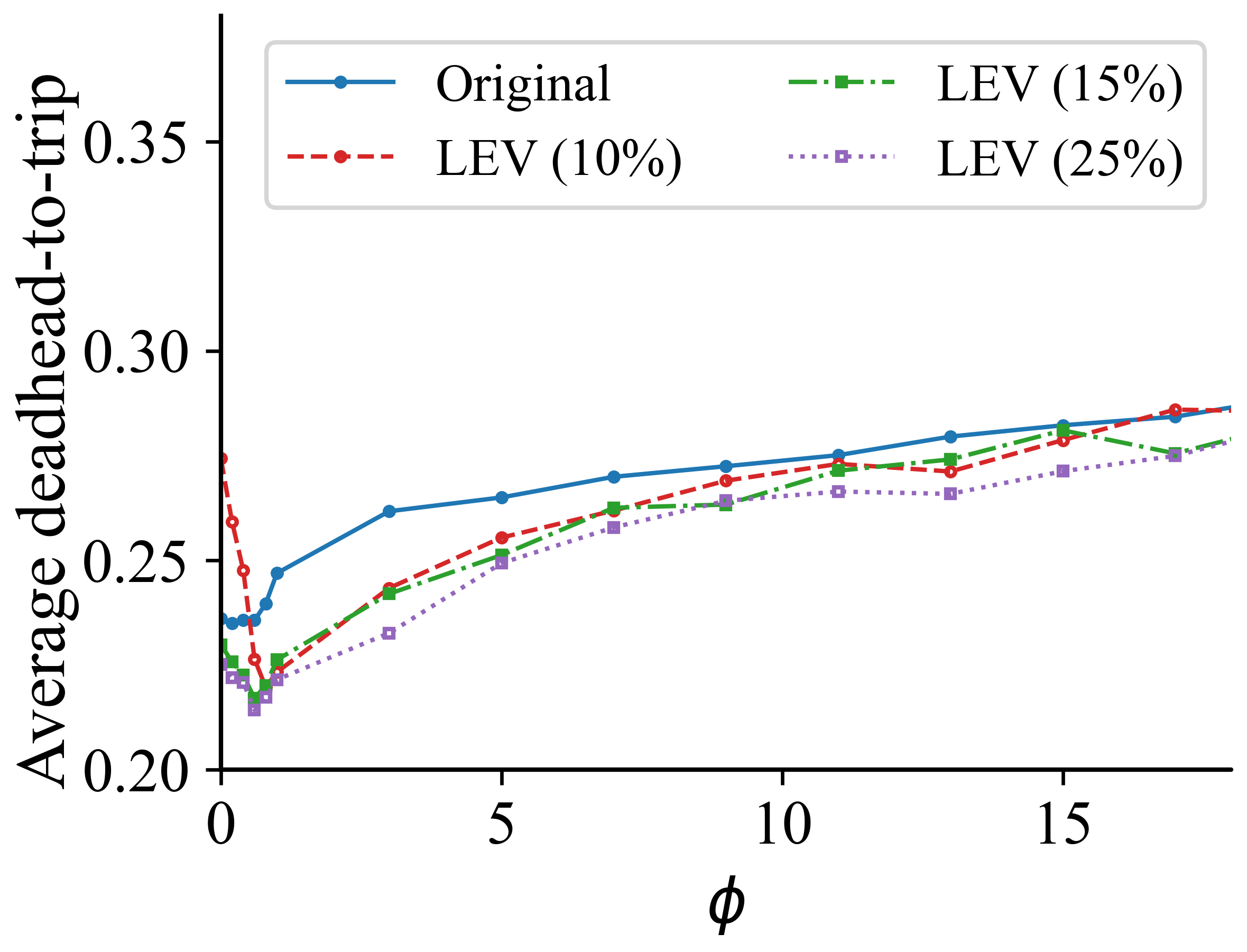} 
  \\
  (a) LEVs &  (b) HEVs
  \end{tabular}
  \caption{
  Average deadhead-to-trip distance ratio on $y$-axis for (a) LEVs and (b) HEVs, as a function of $\phi$ on the $x$-axis for different LEV fractions. At higher $\phi$ values, the deadhead-to-distance ratio for LEVs decreases but increases for HEVs.}
  \label{fig:equity_deadhead}
\end{figure}

A distinct facet of our investigation involved analyzing the ratio between deadhead and total trip distance, known as the deadhead-to-trip ratio. Figure~\ref{fig:equity_deadhead} showcases the results as the threshold values vary for LEVs and HEVs. As expected, an increase in the threshold value is associated with a decrease in the deadhead-to-trip ratio, consistent across scenarios with a fixed percentage of LEVs.

Interestingly, our results demonstrate that raising the threshold for a fixed percentage of LEVs is linked to an increased deadhead-to-trip ratio for HEVs. Additionally, for a fixed threshold value, a higher percentage of LEVs results in a lower deadhead-to-trip ratio, reflecting improved efficiency.  These observations emphasize the intricate relationship between ride assignment parameters and the efficiency of trip assignments in diverse environmental contexts. Results show that the deadhead-to-trip ratio for LEVs varies between $31.8\%$ to $35.7\%$ for original vehicles when threshold varies between $0.001$ to $18$ while these ranges for datasets with 10\%, 15\%, and 25\% LEVs were $32.8\%$ to $37.0\%$, $32.4\%$ to $34.0\%$, and $31.9\%$ to $34\%$ respectively.

\vspace{0.05cm}
\noindent \emph{\textbf{Key Takeaway 2.} Under the ride assignment of \TORA, reduction in deadhead emissions requires sacrificing equity among drivers. In this case, a greater fraction of passengers get assigned to LEVs, and their deadhead miles would be longer.}

\begin{figure*}[t]
    \centering
    \begin{tabular}{cccc}
    \includegraphics[width=0.23 \linewidth, height=0.2 \linewidth]{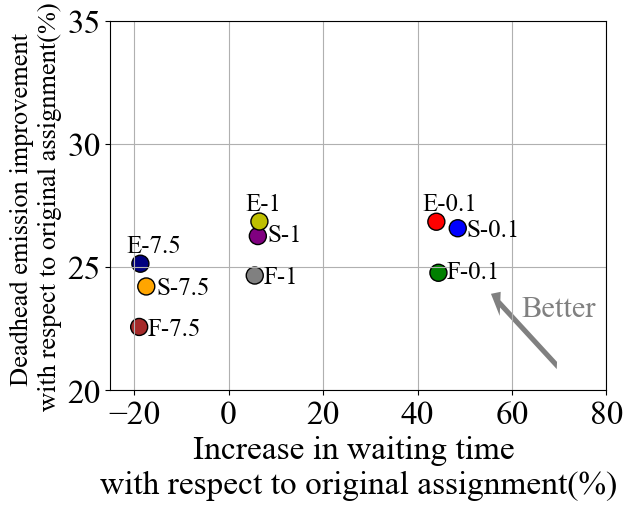} &
    \includegraphics[width=0.23 \linewidth, height=0.2 \linewidth]{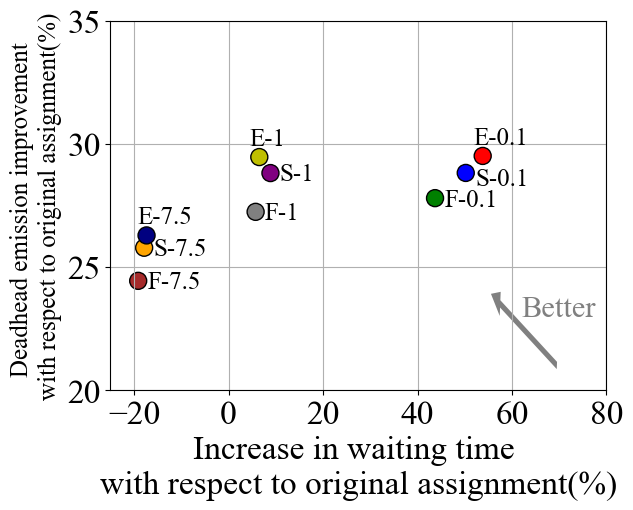} & \includegraphics[width=0.23 \linewidth, height=0.2 \linewidth]{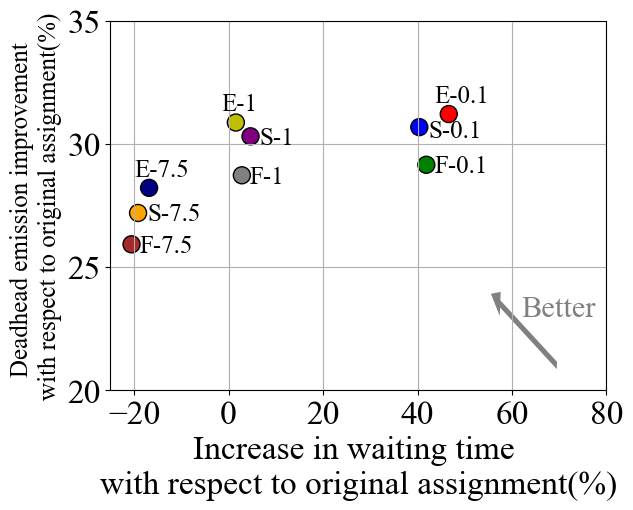} &
    \includegraphics[width=0.23 \linewidth, height=0.2 \linewidth]{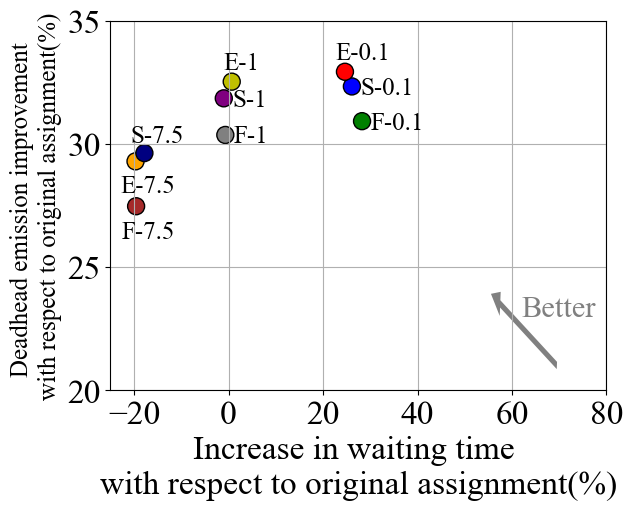}\\
    (a) Original Vehicles  &  (b) $10\%$ LEVs  & (c) $15\%$ LEVs & (d) $25\%$ LEVs
    \end{tabular}
    \caption{Percentage of improvement in total emission  as a function of percentage of increase in waiting time assignment for \TORA using three different route options (Shortest (S), Fastest (F), and Fuel Efficient (E)) and varying $\phi$ values: 0.1, 1, and 7.5.}
    \vspace{-10pt}
    \label{fig:deadhead_vs_waiting} 
\end{figure*}

\subsection{Ride Assignment and Routing Comparison}

Finally, we examine the impact of ride assignments and routing options on total emissions and waiting times, considering datasets with varying percentages of LEVs. By leveraging data from the Google Maps API, including travel times for different route options, the shortest (S), the fastest (F), and the fuel-efficient (E), explained in Section~\ref{sec:routing_tradeOff}, we evaluated the performance of \TORA with three different values of $\phi:$ $0.1$, $1$, and $7.5$. Our results, illustrated in Figure~\ref{fig:deadhead_vs_waiting}, indicate when we increase the threshold value, we observe a decrease in the percentage increase in waiting time. Higher thresholds are associated with a reduced percentage increase in waiting times and a corresponding decrease in improvements in deadhead emissions. Conversely, lowering the threshold value amplifies waiting time increases and enhances the reduction of deadhead emissions. These findings illuminate the intricate dynamics between ride assignment and routing strategies and their consequential effects on emissions and waiting times within ridesharing systems. In addition, results reveal that the impact of ride assignment (different threshold values) on the total emission and waiting times is substantially greater than that of routing algorithms. Note with original vehicles, changing the threshold value from 7.5 to 0.1 can improve the total emission up to $3.5$\%, and increase the waiting time up to $71.2\%$. In contrast, for a fixed ride assignment algorithm, the maximum impact of the routing algorithm on the improvement of emission and increase in waiting time is $2.6\%$, and $10.1\%$. 

\noindent \emph{\textbf{Key Takeaway 3.} The impact of ride assignment algorithms on the emission and waiting time of ride sharing service is significantly more than the impact of routing algorithms.}

%% file: 7-related.tex
 Prior work on improving ride-sharing services has mainly focused on investigating the factors affecting passenger demand and scaling drivers' availability to match demand by leveraging either theoretical approaches or empirical and ML-based approaches.

\vspace{0.05cm}
 \noindent\textbf{Theoretical approaches.}
 Feng et al.~\cite{feng2021two} present a novel two-stage stochastic matching model for ride-hailing platforms, addressing uncertainty in rider and driver availability.
Li et al.~\cite{li2011hunting} study the strategies taken by taxi drivers for finding passengers from a large-scale taxi GPS dataset. 
Vazifeh et al.~\cite{vazifeh2018addressing} propose an optimization-based approach to minimize the fleet size in on-demand urban mobility services. 
Zha et al.~\cite{zha2016economic} analyze the ride-sourcing market by leveraging a model that captures the matchings between customers and drivers through a matching function. Bai et al.~\cite{bai2019coordinating} and Feng et al.~\cite{feng2021we} propose a model for an on-demand service platform to estimate the passenger’s queuing time or their matching time. 
Abkarian et al.~\cite{abkarian2022modeling} present a model that aims to balance the trade-off between waiting times for reservation-based and on-demand users while minimizing the overall deadhead mileage driven by the vehicles in the fleet.
Other studies analyze ride-sharing for deadheading, cost, energy consumption, and environmental impact through modeling, simulations, and data analytics~\cite{bauer2018cost, nair2020model, wenzel2019travel}. 

\vspace{0.05cm}
\noindent\textbf{Empirical and ML-based approaches.} 
Moreira-Matias et al.~\cite{moreira2013predicting} devise a method to predict the short-term distribution of taxi passengers using streaming data from taxis operating in Porto, Portugal. 
Iacobucci et al.~\cite{iacobucci2023demand} investigate the potential demand for Shared Autonomous Vehicles (SAVs) or robotaxis using a scalable simulation framework. This study explores the impact of SAVs on travel behavior, considering factors such as fare, waiting times, and real-time demand. 
Jungel et al. \cite{jungel2023learning} focus on the development of online control algorithms for autonomous mobility-on-demand systems. They propose a hybrid combinatorial optimization enriched machine learning pipeline that learns online dispatching and rebalancing policies from optimal full-information solutions.

Lavieri et al.~\cite{lavieri2018model} use an Austin-based ride-sharing dataset to present two statistical models: (i) a spatially lagged multivariate count model to estimate the number of trips generated in a specific zone during weekdays and weekends and (ii) fractional split model to identify the key features of zones where a majority of ride-sharing trips start. In another study, 
Liu et al.~\cite{liu2017short} analyze the temporal and spatial patterns of riders' requests and the ride services they choose. They use Random Forests to predict the travel demand of these ride-sharing services. 
Ke et al.~\cite{ke2017short} propose a novel spatio-temporal deep learning approach that uses a convolutional neural network (CNN) to model the spatial distribution of demand and a long short-term memory (LSTM) network to model the temporal patterns in ride demand. 
Other studies have also used LSTM networks to predict demand behaviors across both spatial and temporal dimensions~\cite{xu2017real, ku2021real, wang2020data}. While these studies focus on improving the performance of ride-sharing services, they do not explicitly target reducing deadhead miles.
To the best of our knowledge, the most relevant work to our proposal is done by Kontou et al.~\cite{kontou2020reducing}. Authors show up to 82\% reduction in \emph{trip-level} deadhead miles by leveraging hour-ahead trip demand predictions and a heuristic approach to driver assignment. 
However, their focus on reducing trip-level deadhead miles may not always lead to reducing system-wide deadhead miles and emissions that, in addition to the number of miles, depend on the fuel efficiency of vehicles and traffic conditions. 
Furthermore, they do not consider equity metrics from the rider's or driver's perspective.

Our paper is the first to take a holistic approach toward designing emission-aware ride assignment optimizations that explicitly target reducing emissions from deadhead miles, embed equity considerations into the ride assignment process, and consider electric and low-emission vehicles as a part of ride-sharing fleets. In addition, our solution approach is a data-driven algorithmic design that relies on a straightforward greedy algorithm. 

Lastly, we note that there is extensive literature on optimizing carbon emissions and energy consumption for vehicle routing. The existing emission-optimized routing algorithms, e.g.,~\cite{deng2016energy, liu2018energy, xu2023ride}, mainly model the problem as extended variants of the classic traveling salesman problem under different settings, e.g., joint electric vehicle charging and routing~\cite{desaulniers2016exact, goeke2015routing, montoya2017electric, schneider2014electric}, joint routing and speed optimization~\cite{su2023follow, deng2016energy}, and environmental impacts of the routing decisions \cite{erdougan2012green, bektacs2011pollution}.
In this work, however, we used off-the-shelf routing algorithms from Google Map API. An interesting future direction of our work is to combine the above routing algorithms with the ride assignment policies presented in this paper.

%% file: 8-conclusion.tex
In this paper, we presented equity- and emission-aware ride assignment and routing approaches to reduce the overall emissions of the ridesharing platform. We then presented \kit as a toolkit that combines multiple datasets needed for the performance evaluation of our algorithms.
Future research should focus on explicitly integrating equity considerations into the ride assignment optimization problem. This involves developing models that account for various dimensions of equity, such as fairness in service access and drivers' treatment, to align the optimization process with broader social goals. Another critical area for further exploration is a thorough analysis of the simple online algorithm used in this study. Investigating its robustness, scalability, and sensitivity to different conditions will provide valuable insights, enabling potential refinements and improvements to enhance its performance under various scenarios.
Additionally, beyond carbon minimization, future research should extend the evaluation framework to include a comprehensive assessment of the impact on drivers' income. Balancing environmental objectives with socioeconomic considerations is essential to ensure that emission reduction efforts do not compromise the livelihoods of those providing ride-sharing services.
Finally, we will publicly release \kit as an open-source toolkit to facilitate research in emissions analysis of ridesharing ecosystems.

%% file: 9-ack.tex
This research was supported in part by NSF grants CAREER 2045641, CNS-2102963, CNS-2106299, CPS-2136199, NRT-2021693, and NGSDI-2105494.

%% file: paper.bbl

\begin{thebibliography}{61}


\ifx \showCODEN    \undefined \def \showCODEN     #1{\unskip}     \fi
\ifx \showDOI      \undefined \def \showDOI       #1{#1}\fi
\ifx \showISBNx    \undefined \def \showISBNx     #1{\unskip}     \fi
\ifx \showISBNxiii \undefined \def \showISBNxiii  #1{\unskip}     \fi
\ifx \showISSN     \undefined \def \showISSN      #1{\unskip}     \fi
\ifx \showLCCN     \undefined \def \showLCCN      #1{\unskip}     \fi
\ifx \shownote     \undefined \def \shownote      #1{#1}          \fi
\ifx \showarticletitle \undefined \def \showarticletitle #1{#1}   \fi
\ifx \showURL      \undefined \def \showURL       {\relax}        \fi
\providecommand\bibfield[2]{#2}
\providecommand\bibinfo[2]{#2}
\providecommand\natexlab[1]{#1}
\providecommand\showeprint[2][]{arXiv:#2}

\bibitem[ele(2020)]%
        {electricity-map}
 \bibinfo{year}{2020}\natexlab{}.
\newblock \bibinfo{title}{Electricity {M}ap}.
\newblock \bibinfo{howpublished}{\url{https://www.electricitymap.org/map}}.
\newblock


\bibitem[reu(2020)]%
        {reuters}
 \bibinfo{year}{2020}\natexlab{}.
\newblock \bibinfo{title}{Ride-hailing increases emissions, contributes to climate pollution - study}.
\newblock \bibinfo{howpublished}{\url{https://www.reuters.com/article/us-uber-emissions/ride-hailing-increases-emissions-contributes-to-climate-pollution-study-idUSKBN20J27K}}.
\newblock
\newblock
\shownote{(Accessed on 05/05/2020)}.


\bibitem[geo(2023)]%
        {geocodIO}
 \bibinfo{year}{2023}\natexlab{}.
\newblock \bibinfo{title}{Geocodio's RESTful API}.
\newblock \bibinfo{howpublished}{\url{https://www.geocod.io/docs/}}.
\newblock


\bibitem[wat(2023)]%
        {watttime}
 \bibinfo{year}{2023}\natexlab{}.
\newblock \bibinfo{title}{How clean is the electricity {I}’m using right now?}
\newblock \bibinfo{howpublished}{\url{https://www.watttime.org/}}.
\newblock
\newblock
\shownote{Accessed: 2023-02-27}.


\bibitem[blo(2023)]%
        {bloombergRideSharing}
 \bibinfo{year}{2023}\natexlab{}.
\newblock \bibinfo{title}{Ride Sharing Market Size to Hit USD 205.83 Billion, Globally, by the end of 2030, Growing at a CAGR of 13.5\%}.
\newblock \bibinfo{howpublished}{\url{https://tinyurl.com/BloombergRideSharing}}.
\newblock
\newblock
\shownote{Access: 2023-05-20}.


\bibitem[Abkarian et~al\mbox{.}(2022)]%
        {abkarian2022modeling}
\bibfield{author}{\bibinfo{person}{Hoseb Abkarian}, \bibinfo{person}{Hani~S Mahmassani}, {and} \bibinfo{person}{Michael Hyland}.} \bibinfo{year}{2022}\natexlab{}.
\newblock \showarticletitle{Modeling the mixed-service fleet problem of shared-use autonomous mobility systems for on-demand ridesourcing and carsharing with reservations}.
\newblock \bibinfo{journal}{\emph{Transportation Research Record}} \bibinfo{volume}{2676}, \bibinfo{number}{8} (\bibinfo{year}{2022}), \bibinfo{pages}{363--375}.
\newblock


\bibitem[Anair et~al\mbox{.}(2020)]%
        {ucsusa}
\bibfield{author}{\bibinfo{person}{Don Anair}, \bibinfo{person}{Jeremy Martin}, \bibinfo{person}{Maria Cecilia~Pinto de Moura}, {and} \bibinfo{person}{Joshua Goldman}.} \bibinfo{year}{2020}\natexlab{}.
\newblock \bibinfo{title}{Ride-Hailing's Climate Risks}.
\newblock \bibinfo{howpublished}{\url{https://www.ucsusa.org/resources/ride-hailing-climate-risks}}.
\newblock
\newblock
\shownote{(Accessed on 05/05/2020)}.


\bibitem[Bai et~al\mbox{.}(2019)]%
        {bai2019coordinating}
\bibfield{author}{\bibinfo{person}{Jiaru Bai}, \bibinfo{person}{Kut~C So}, \bibinfo{person}{Christopher~S Tang}, \bibinfo{person}{Xiqun Chen}, {and} \bibinfo{person}{Hai Wang}.} \bibinfo{year}{2019}\natexlab{}.
\newblock \showarticletitle{Coordinating supply and demand on an on-demand service platform with impatient customers}.
\newblock \bibinfo{journal}{\emph{Manufacturing \& Service Operations Management}} \bibinfo{volume}{21}, \bibinfo{number}{3} (\bibinfo{year}{2019}), \bibinfo{pages}{556--570}.
\newblock


\bibitem[Bauer et~al\mbox{.}(2018)]%
        {bauer2018cost}
\bibfield{author}{\bibinfo{person}{Gordon~S Bauer}, \bibinfo{person}{Jeffery~B Greenblatt}, {and} \bibinfo{person}{Brian~F Gerke}.} \bibinfo{year}{2018}\natexlab{}.
\newblock \showarticletitle{Cost, energy, and environmental impact of automated electric taxi fleets in {M}anhattan}.
\newblock \bibinfo{journal}{\emph{Environmental science \& technology}} \bibinfo{volume}{52}, \bibinfo{number}{8} (\bibinfo{year}{2018}), \bibinfo{pages}{4920--4928}.
\newblock


\bibitem[Bekta{\c{s}} and Laporte(2011)]%
        {bektacs2011pollution}
\bibfield{author}{\bibinfo{person}{Tolga Bekta{\c{s}}} {and} \bibinfo{person}{Gilbert Laporte}.} \bibinfo{year}{2011}\natexlab{}.
\newblock \showarticletitle{The pollution-routing problem}.
\newblock \bibinfo{journal}{\emph{Transportation Research Part B: Methodological}} \bibinfo{volume}{45}, \bibinfo{number}{8} (\bibinfo{year}{2011}), \bibinfo{pages}{1232--1250}.
\newblock


\bibitem[Bliss(2019)]%
        {citylab}
\bibfield{author}{\bibinfo{person}{Laura Bliss}.} \bibinfo{year}{2019}\natexlab{}.
\newblock \bibinfo{title}{How Much Traffic Do {U}ber and {L}yft Cause?}
\newblock \bibinfo{howpublished}{\url{https://www.citylab.com/transportation/2019/08/uber-lyft-traffic-congestion-ride-hailing-cities-drivers-vmt/595393/}}.
\newblock


\bibitem[Bureau(2020)]%
        {Census:2020}
\bibfield{author}{\bibinfo{person}{United States~Census Bureau}.} \bibinfo{year}{2020}\natexlab{}.
\newblock \bibinfo{title}{Census Data}.
\newblock
\newblock
\urldef\tempurl%
\url{https://data.census.gov/cedsci/}
\showURL{%
\tempurl}


\bibitem[{Congressional Budget Office}(2022)]%
        {cbo2022emissions}
\bibfield{author}{\bibinfo{person}{{Congressional Budget Office}}.} \bibinfo{year}{2022}\natexlab{}.
\newblock \bibinfo{title}{Emissions of Carbon Dioxide in the Transportation Sector}.
\newblock \bibinfo{howpublished}{\url{https://www.cbo.gov/publication/58566}}.
\newblock


\bibitem[Cook et~al\mbox{.}(2021)]%
        {cook2021gender}
\bibfield{author}{\bibinfo{person}{Cody Cook}, \bibinfo{person}{Rebecca Diamond}, \bibinfo{person}{Jonathan~V Hall}, \bibinfo{person}{John~A List}, {and} \bibinfo{person}{Paul Oyer}.} \bibinfo{year}{2021}\natexlab{}.
\newblock \showarticletitle{The gender earnings gap in the gig economy: Evidence from over a million rideshare drivers}.
\newblock \bibinfo{journal}{\emph{The Review of Economic Studies}} \bibinfo{volume}{88}, \bibinfo{number}{5} (\bibinfo{year}{2021}), \bibinfo{pages}{2210--2238}.
\newblock


\bibitem[Cramer and Krueger(2016)]%
        {cramer2016disruptive}
\bibfield{author}{\bibinfo{person}{Judd Cramer} {and} \bibinfo{person}{Alan~B Krueger}.} \bibinfo{year}{2016}\natexlab{}.
\newblock \showarticletitle{Disruptive change in the taxi business: The case of Uber}.
\newblock \bibinfo{journal}{\emph{American Economic Review}} \bibinfo{volume}{106}, \bibinfo{number}{5} (\bibinfo{year}{2016}), \bibinfo{pages}{177--182}.
\newblock


\bibitem[Deng et~al\mbox{.}(2016)]%
        {deng2016energy}
\bibfield{author}{\bibinfo{person}{Lei Deng}, \bibinfo{person}{Mohammad~H. Hajiesmaili}, \bibinfo{person}{Minghua Chen}, {and} \bibinfo{person}{Haibo Zeng}.} \bibinfo{year}{2016}\natexlab{}.
\newblock \showarticletitle{Energy-efficient Timely Transportation of Long-haul Heavy-duty Trucks}. In \bibinfo{booktitle}{\emph{Proc. of ACM International Conference on Future Energy Systems (eEnergy)}}.
\newblock


\bibitem[Desaulniers et~al\mbox{.}(2016)]%
        {desaulniers2016exact}
\bibfield{author}{\bibinfo{person}{Guy Desaulniers}, \bibinfo{person}{Fausto Errico}, \bibinfo{person}{Stefan Irnich}, {and} \bibinfo{person}{Michael Schneider}.} \bibinfo{year}{2016}\natexlab{}.
\newblock \showarticletitle{Exact algorithms for electric vehicle-routing problems with time windows}.
\newblock \bibinfo{journal}{\emph{Operations Research}} \bibinfo{volume}{64}, \bibinfo{number}{6} (\bibinfo{year}{2016}), \bibinfo{pages}{1388--1405}.
\newblock


\bibitem[{Directorate-General for Mobility and Transport, European Commission}(2023)]%
        {eu2023urban}
\bibfield{author}{\bibinfo{person}{{Directorate-General for Mobility and Transport, European Commission}}.} \bibinfo{year}{2023}\natexlab{}.
\newblock \bibinfo{title}{Urban mobility}.
\newblock \bibinfo{howpublished}{\url{https://transport.ec.europa.eu/transport-themes/clean-transport-urban-transport/urban-mobility_en}}.
\newblock


\bibitem[Erdo{\u{g}}an and Miller-Hooks(2012)]%
        {erdougan2012green}
\bibfield{author}{\bibinfo{person}{Sevgi Erdo{\u{g}}an} {and} \bibinfo{person}{Elise Miller-Hooks}.} \bibinfo{year}{2012}\natexlab{}.
\newblock \showarticletitle{A green vehicle routing problem}.
\newblock \bibinfo{journal}{\emph{Transportation research part E: logistics and transportation review}} \bibinfo{volume}{48}, \bibinfo{number}{1} (\bibinfo{year}{2012}), \bibinfo{pages}{100--114}.
\newblock


\bibitem[Feng et~al\mbox{.}(2021a)]%
        {feng2021we}
\bibfield{author}{\bibinfo{person}{Guiyun Feng}, \bibinfo{person}{Guangwen Kong}, {and} \bibinfo{person}{Zizhuo Wang}.} \bibinfo{year}{2021}\natexlab{a}.
\newblock \showarticletitle{We are on the way: Analysis of on-demand ride-hailing systems}.
\newblock \bibinfo{journal}{\emph{Manufacturing \& Service Operations Management}} \bibinfo{volume}{23}, \bibinfo{number}{5} (\bibinfo{year}{2021}), \bibinfo{pages}{1237--1256}.
\newblock


\bibitem[Feng et~al\mbox{.}(2021b)]%
        {feng2021two}
\bibfield{author}{\bibinfo{person}{Yiding Feng}, \bibinfo{person}{Rad Niazadeh}, {and} \bibinfo{person}{Amin Saberi}.} \bibinfo{year}{2021}\natexlab{b}.
\newblock \showarticletitle{Two-stage stochastic matching with application to ride hailing}. In \bibinfo{booktitle}{\emph{Proceedings of the 2021 ACM-SIAM Symposium on Discrete Algorithms (SODA)}}. SIAM, \bibinfo{pages}{2862--2877}.
\newblock


\bibitem[Fielbaum(2022)]%
        {fielbaum2022optimizing}
\bibfield{author}{\bibinfo{person}{Andr{\'e}s Fielbaum}.} \bibinfo{year}{2022}\natexlab{}.
\newblock \showarticletitle{Optimizing a vehicle’s route in an on-demand ridesharing system in which users might walk}.
\newblock \bibinfo{journal}{\emph{Journal of Intelligent Transportation Systems}} \bibinfo{volume}{26}, \bibinfo{number}{4} (\bibinfo{year}{2022}), \bibinfo{pages}{432--447}.
\newblock


\bibitem[Goeke and Schneider(2015)]%
        {goeke2015routing}
\bibfield{author}{\bibinfo{person}{Dominik Goeke} {and} \bibinfo{person}{Michael Schneider}.} \bibinfo{year}{2015}\natexlab{}.
\newblock \showarticletitle{Routing a mixed fleet of electric and conventional vehicles}.
\newblock \bibinfo{journal}{\emph{European Journal of Operational Research}} \bibinfo{volume}{245}, \bibinfo{number}{1} (\bibinfo{year}{2015}), \bibinfo{pages}{81--99}.
\newblock


\bibitem[Google(2023)]%
        {gmapapi}
\bibfield{author}{\bibinfo{person}{Google}.} \bibinfo{year}{2023}\natexlab{}.
\newblock \bibinfo{title}{gmapapi}.
\newblock \bibinfo{howpublished}{\url{https://developers.google.com/maps}}.
\newblock
\newblock
\shownote{Accessed: 2023-09}.


\bibitem[Graham and Shaw(2017)]%
        {graham2017towards}
\bibfield{author}{\bibinfo{person}{Mark Graham} {and} \bibinfo{person}{Joe Shaw}.} \bibinfo{year}{2017}\natexlab{}.
\newblock \bibinfo{booktitle}{\emph{Towards a fairer gig economy}}.
\newblock \bibinfo{publisher}{Meatspace Press}.
\newblock


\bibitem[Henao(2017)]%
        {Henao2017}
\bibfield{author}{\bibinfo{person}{Alejandro Henao}.} \bibinfo{year}{2017}\natexlab{}.
\newblock \bibinfo{booktitle}{\emph{Impacts of Ridesourcing-Lyft and Uber-on Transportation Including VMT, Mode Replacement, Parking, and Travel Behavior}}.
\newblock \bibinfo{publisher}{University of Colorado at Denver}.
\newblock


\bibitem[Henao and Marshall(2019)]%
        {henao2019impact}
\bibfield{author}{\bibinfo{person}{Alejandro Henao} {and} \bibinfo{person}{Wesley~E Marshall}.} \bibinfo{year}{2019}\natexlab{}.
\newblock \showarticletitle{The impact of ride-hailing on vehicle miles traveled}.
\newblock \bibinfo{journal}{\emph{Transportation}} \bibinfo{volume}{46}, \bibinfo{number}{6} (\bibinfo{year}{2019}), \bibinfo{pages}{2173--2194}.
\newblock


\bibitem[Iacobucci et~al\mbox{.}(2023)]%
        {iacobucci2023demand}
\bibfield{author}{\bibinfo{person}{Riccardo Iacobucci}, \bibinfo{person}{Jonas Donhauser}, \bibinfo{person}{Jan-Dirk Schm{\"o}cker}, {and} \bibinfo{person}{Marco Pruckner}.} \bibinfo{year}{2023}\natexlab{}.
\newblock \showarticletitle{The demand potential of shared autonomous vehicles: a large-scale simulation using mobility survey data}.
\newblock \bibinfo{journal}{\emph{Journal of Intelligent Transportation Systems}} (\bibinfo{year}{2023}), \bibinfo{pages}{1--22}.
\newblock


\bibitem[{International Energy Agency}(2022)]%
        {iea2022world}
\bibfield{author}{\bibinfo{person}{{International Energy Agency}}.} \bibinfo{year}{2022}\natexlab{}.
\newblock \bibinfo{title}{World Energy Outlook 2022}.
\newblock \bibinfo{howpublished}{\url{https://www.iea.org/reports/world-energy-outlook-2022}}.
\newblock


\bibitem[Jiang et~al\mbox{.}(2018)]%
        {jiang2018ridesharing}
\bibfield{author}{\bibinfo{person}{Shan Jiang}, \bibinfo{person}{Le Chen}, \bibinfo{person}{Alan Mislove}, {and} \bibinfo{person}{Christo Wilson}.} \bibinfo{year}{2018}\natexlab{}.
\newblock \showarticletitle{On ridesharing competition and accessibility: Evidence from {U}ber, {L}yft, and taxi}. In \bibinfo{booktitle}{\emph{Proceedings of the 2018 World Wide Web Conference}}. \bibinfo{pages}{863--872}.
\newblock


\bibitem[Jungel et~al\mbox{.}(2023)]%
        {jungel2023learning}
\bibfield{author}{\bibinfo{person}{Kai Jungel}, \bibinfo{person}{Axel Parmentier}, \bibinfo{person}{Maximilian Schiffer}, {and} \bibinfo{person}{Thibaut Vidal}.} \bibinfo{year}{2023}\natexlab{}.
\newblock \showarticletitle{Learning-based Online Optimization for Autonomous Mobility-on-Demand Fleet Control}.
\newblock \bibinfo{journal}{\emph{arXiv preprint arXiv:2302.03963}} (\bibinfo{year}{2023}).
\newblock


\bibitem[Ke et~al\mbox{.}(2017)]%
        {ke2017short}
\bibfield{author}{\bibinfo{person}{Jintao Ke}, \bibinfo{person}{Hongyu Zheng}, \bibinfo{person}{Hai Yang}, {and} \bibinfo{person}{Xiqun~Michael Chen}.} \bibinfo{year}{2017}\natexlab{}.
\newblock \showarticletitle{Short-term forecasting of passenger demand under on-demand ride services: A spatio-temporal deep learning approach}.
\newblock \bibinfo{journal}{\emph{Transportation research part C: Emerging technologies}}  \bibinfo{volume}{85} (\bibinfo{year}{2017}), \bibinfo{pages}{591--608}.
\newblock


\bibitem[{Knowledge at Wharton Staff}(2020)]%
        {upenn}
\bibfield{author}{\bibinfo{person}{{Knowledge at Wharton Staff}}.} \bibinfo{year}{2020}\natexlab{}.
\newblock \bibinfo{title}{How Green is the Sharing Economy?}
\newblock \bibinfo{howpublished}{\url{https://knowledge.wharton.upenn.edu/article/how-green-is-the-sharing-economy/}}.
\newblock
\newblock
\shownote{(Accessed on 05/05/2020)}.


\bibitem[Komanduri et~al\mbox{.}(2018)]%
        {komanduri2018assessing}
\bibfield{author}{\bibinfo{person}{Anurag Komanduri}, \bibinfo{person}{Zeina Wafa}, \bibinfo{person}{Kimon Proussaloglou}, {and} \bibinfo{person}{Simon Jacobs}.} \bibinfo{year}{2018}\natexlab{}.
\newblock \showarticletitle{Assessing the impact of app-based ride share systems in an urban context: Findings from {A}ustin}.
\newblock \bibinfo{journal}{\emph{Transportation Research Record}} \bibinfo{volume}{2672}, \bibinfo{number}{7} (\bibinfo{year}{2018}), \bibinfo{pages}{34--46}.
\newblock


\bibitem[Kontou et~al\mbox{.}(2020)]%
        {kontou2020reducing}
\bibfield{author}{\bibinfo{person}{Eleftheria Kontou}, \bibinfo{person}{Venu Garikapati}, {and} \bibinfo{person}{Yi Hou}.} \bibinfo{year}{2020}\natexlab{}.
\newblock \showarticletitle{Reducing ridesourcing empty vehicle travel with future travel demand prediction}.
\newblock   \bibinfo{volume}{121} (\bibinfo{year}{2020}), \bibinfo{pages}{102826}.
\newblock
\showISSN{0968-090X}
\urldef\tempurl%
\url{https://doi.org/10.1016/j.trc.2020.102826}
\showDOI{\tempurl}


\bibitem[Ku et~al\mbox{.}(2021)]%
        {ku2021real}
\bibfield{author}{\bibinfo{person}{Donggyun Ku}, \bibinfo{person}{Sungyong Na}, \bibinfo{person}{Jooyoung Kim}, {and} \bibinfo{person}{Seungjae Lee}.} \bibinfo{year}{2021}\natexlab{}.
\newblock \showarticletitle{Real-time taxi demand prediction using recurrent neural network}. In \bibinfo{booktitle}{\emph{Proceedings of the Institution of Civil Engineers-Municipal Engineer}}, Vol.~\bibinfo{volume}{174}. Thomas Telford Ltd, \bibinfo{pages}{75--87}.
\newblock


\bibitem[Lavieri et~al\mbox{.}(2018)]%
        {lavieri2018model}
\bibfield{author}{\bibinfo{person}{Patr{\'\i}cia~S Lavieri}, \bibinfo{person}{Felipe~F Dias}, \bibinfo{person}{Natalia~Ruiz Juri}, \bibinfo{person}{James Kuhr}, {and} \bibinfo{person}{Chandra~R Bhat}.} \bibinfo{year}{2018}\natexlab{}.
\newblock \showarticletitle{A model of ridesourcing demand generation and distribution}.
\newblock \bibinfo{journal}{\emph{Transportation Research Record}} \bibinfo{volume}{2672}, \bibinfo{number}{46} (\bibinfo{year}{2018}), \bibinfo{pages}{31--40}.
\newblock


\bibitem[Li et~al\mbox{.}(2011)]%
        {li2011hunting}
\bibfield{author}{\bibinfo{person}{Bin Li}, \bibinfo{person}{Daqing Zhang}, \bibinfo{person}{Lin Sun}, \bibinfo{person}{Chao Chen}, \bibinfo{person}{Shijian Li}, \bibinfo{person}{Guande Qi}, {and} \bibinfo{person}{Qiang Yang}.} \bibinfo{year}{2011}\natexlab{}.
\newblock \showarticletitle{Hunting or waiting? Discovering passenger-finding strategies from a large-scale real-world taxi dataset}. In \bibinfo{booktitle}{\emph{2011 IEEE International Conference on Pervasive Computing and Communications Workshops (PERCOM Workshops)}}. IEEE, \bibinfo{pages}{63--68}.
\newblock


\bibitem[Liu et~al\mbox{.}(2017)]%
        {liu2017short}
\bibfield{author}{\bibinfo{person}{Jiaokun Liu}, \bibinfo{person}{Erjia Cui}, \bibinfo{person}{Haoqiang Hu}, \bibinfo{person}{Xiaowei Chen}, \bibinfo{person}{Xiqun Chen}, {and} \bibinfo{person}{Feng Chen}.} \bibinfo{year}{2017}\natexlab{}.
\newblock \showarticletitle{Short-term forecasting of emerging on-demand ride services}. In \bibinfo{booktitle}{\emph{2017 4th International Conference on Transportation Information and Safety (ICTIS)}}. IEEE, \bibinfo{pages}{489--495}.
\newblock


\bibitem[Liu et~al\mbox{.}(2018)]%
        {liu2018energy}
\bibfield{author}{\bibinfo{person}{Qingyu Liu}, \bibinfo{person}{Haibo Zeng}, {and} \bibinfo{person}{Minghua Chen}.} \bibinfo{year}{2018}\natexlab{}.
\newblock \showarticletitle{Energy-efficient timely truck transportation for geographically-dispersed tasks}. In \bibinfo{booktitle}{\emph{Proceedings of the Ninth International Conference on Future Energy Systems}}. \bibinfo{pages}{324--339}.
\newblock


\bibitem[Lloyd(1982)]%
        {lloyd1982least}
\bibfield{author}{\bibinfo{person}{Stuart Lloyd}.} \bibinfo{year}{1982}\natexlab{}.
\newblock \showarticletitle{Least {S}quares {Q}uantization in {P}{C}{M}}.
\newblock \bibinfo{journal}{\emph{IEEE Transactions on Information Theory}} (\bibinfo{year}{1982}).
\newblock


\bibitem[Montoya et~al\mbox{.}(2017)]%
        {montoya2017electric}
\bibfield{author}{\bibinfo{person}{Alejandro Montoya}, \bibinfo{person}{Christelle Gu{\'e}ret}, \bibinfo{person}{Jorge~E Mendoza}, {and} \bibinfo{person}{Juan~G Villegas}.} \bibinfo{year}{2017}\natexlab{}.
\newblock \showarticletitle{The electric vehicle routing problem with nonlinear charging function}.
\newblock \bibinfo{journal}{\emph{Transportation Research Part B: Methodological}}  \bibinfo{volume}{103} (\bibinfo{year}{2017}), \bibinfo{pages}{87--110}.
\newblock


\bibitem[Moreira-Matias et~al\mbox{.}(2013)]%
        {moreira2013predicting}
\bibfield{author}{\bibinfo{person}{Luis Moreira-Matias}, \bibinfo{person}{Joao Gama}, \bibinfo{person}{Michel Ferreira}, \bibinfo{person}{Joao Mendes-Moreira}, {and} \bibinfo{person}{Luis Damas}.} \bibinfo{year}{2013}\natexlab{}.
\newblock \showarticletitle{Predicting taxi--passenger demand using streaming data}.
\newblock \bibinfo{journal}{\emph{IEEE Transactions on Intelligent Transportation Systems}} \bibinfo{volume}{14}, \bibinfo{number}{3} (\bibinfo{year}{2013}), \bibinfo{pages}{1393--1402}.
\newblock


\bibitem[Mourad et~al\mbox{.}(2019)]%
        {mourad2019survey}
\bibfield{author}{\bibinfo{person}{Abood Mourad}, \bibinfo{person}{Jakob Puchinger}, {and} \bibinfo{person}{Chengbin Chu}.} \bibinfo{year}{2019}\natexlab{}.
\newblock \showarticletitle{A survey of models and algorithms for optimizing shared mobility}.
\newblock \bibinfo{journal}{\emph{Transportation Research Part B: Methodological}}  \bibinfo{volume}{123} (\bibinfo{year}{2019}), \bibinfo{pages}{323--346}.
\newblock


\bibitem[Nair et~al\mbox{.}(2020)]%
        {nair2020model}
\bibfield{author}{\bibinfo{person}{Gopindra~S Nair}, \bibinfo{person}{Chandra~R Bhat}, \bibinfo{person}{Irfan Batur}, \bibinfo{person}{Ram~M Pendyala}, {and} \bibinfo{person}{William~HK Lam}.} \bibinfo{year}{2020}\natexlab{}.
\newblock \showarticletitle{A model of deadheading trips and pick-up locations for ride-hailing service vehicles}.
\newblock \bibinfo{journal}{\emph{Transportation Research Part A: Policy and Practice}}  \bibinfo{volume}{135} (\bibinfo{year}{2020}), \bibinfo{pages}{289--308}.
\newblock


\bibitem[of~Energy(2023)]%
        {charging-locations}
\bibfield{author}{\bibinfo{person}{U.S.~Department of Energy}.} \bibinfo{year}{2023}\natexlab{}.
\newblock \bibinfo{title}{Electric Vehicle Charging Station Locations}.
\newblock \bibinfo{howpublished}{\url{https://afdc.energy.gov/fuels/electricity_locations.html\#/find/nearest?fuel=ELEC}}.
\newblock


\bibitem[Oke et~al\mbox{.}(2019)]%
        {oke2019novel}
\bibfield{author}{\bibinfo{person}{Jimi~B. Oke}, \bibinfo{person}{Youssef~M. Aboutaleb}, \bibinfo{person}{Arun Akkinepally}, \bibinfo{person}{Carlos~Lima Azevedo}, \bibinfo{person}{Yafei Han}, \bibinfo{person}{P.~Christopher Zegras}, \bibinfo{person}{Joseph Ferreira}, {and} \bibinfo{person}{Moshe~E. Ben-Akiva}.} \bibinfo{year}{2019}\natexlab{}.
\newblock \showarticletitle{A novel global urban typology framework for sustainable mobility futures}.
\newblock \bibinfo{journal}{\emph{Environmental Research Letters}} \bibinfo{volume}{14}, \bibinfo{number}{9} (\bibinfo{year}{2019}), \bibinfo{pages}{095006}.
\newblock


\bibitem[RideAustin(2017)]%
        {rideaustin-dataset}
\bibfield{author}{\bibinfo{person}{RideAustin}.} \bibinfo{year}{2017}\natexlab{}.
\newblock \bibinfo{title}{Ride-Austin-june6-april13}.
\newblock \bibinfo{howpublished}{\url{https://data.world/ride-austin/ride-austin-june-6-april-13}}.
\newblock


\bibitem[Rosenblat et~al\mbox{.}(2016)]%
        {rosenblat2016discriminating}
\bibfield{author}{\bibinfo{person}{Alex Rosenblat}, \bibinfo{person}{Karen Levy}, \bibinfo{person}{Solon Barocas}, {and} \bibinfo{person}{Tim Hwang}.} \bibinfo{year}{2016}\natexlab{}.
\newblock \showarticletitle{Discriminating tastes: Customer ratings as vehicles for bias}.
\newblock \bibinfo{journal}{\emph{Data \& Society}} (\bibinfo{year}{2016}), \bibinfo{pages}{1--21}.
\newblock


\bibitem[Schaller(2017)]%
        {Schaller2017}
\bibfield{author}{\bibinfo{person}{Bruce Schaller}.} \bibinfo{year}{2017}\natexlab{}.
\newblock \bibinfo{booktitle}{\emph{Unsustainable? {T}he Growth of App-Based Ride Services and Traffic, Travel and the Future of {N}ew {Y}ork {C}ity}}.
\newblock \bibinfo{type}{{T}echnical {R}eport}. \bibinfo{institution}{Schaller Consulting}, \bibinfo{address}{New York, NY}.
\newblock


\bibitem[Schneider et~al\mbox{.}(2014)]%
        {schneider2014electric}
\bibfield{author}{\bibinfo{person}{Michael Schneider}, \bibinfo{person}{Andreas Stenger}, {and} \bibinfo{person}{Dominik Goeke}.} \bibinfo{year}{2014}\natexlab{}.
\newblock \showarticletitle{The electric vehicle-routing problem with time windows and recharging stations}.
\newblock \bibinfo{journal}{\emph{Transportation science}} \bibinfo{volume}{48}, \bibinfo{number}{4} (\bibinfo{year}{2014}), \bibinfo{pages}{500--520}.
\newblock


\bibitem[Su et~al\mbox{.}(2023)]%
        {su2023follow}
\bibfield{author}{\bibinfo{person}{Junyan Su}, \bibinfo{person}{Qiulin Lin}, {and} \bibinfo{person}{Minghua Chen}.} \bibinfo{year}{2023}\natexlab{}.
\newblock \showarticletitle{Follow the Sun and Go with the Wind: Carbon Footprint Optimized Timely E-Truck Transportation}. In \bibinfo{booktitle}{\emph{Proceedings of the 14th ACM International Conference on Future Energy Systems}}. \bibinfo{pages}{159--171}.
\newblock


\bibitem[{U.S. Census Bureau}(2023a)]%
        {us-census}
\bibfield{author}{\bibinfo{person}{{U.S. Census Bureau}}.} \bibinfo{year}{2023}\natexlab{a}.
\newblock \bibinfo{title}{Measuring {A}merica's People, Places, and Economy}.
\newblock \bibinfo{howpublished}{\url{https://www.census.gov/}}.
\newblock


\bibitem[{U.S. Census Bureau}(2023b)]%
        {canadacar_emissions}
\bibfield{author}{\bibinfo{person}{{U.S. Census Bureau}}.} \bibinfo{year}{2023}\natexlab{b}.
\newblock \bibinfo{title}{Vehicle Emissions Forecasting}.
\newblock \bibinfo{howpublished}{\url{https://www.kaggle.com/datasets/abhikdas2809/canadacaremissions}}.
\newblock


\bibitem[Vazifeh et~al\mbox{.}(2018)]%
        {vazifeh2018addressing}
\bibfield{author}{\bibinfo{person}{Mohammad~M Vazifeh}, \bibinfo{person}{Paolo Santi}, \bibinfo{person}{Giovanni Resta}, \bibinfo{person}{Steven~H Strogatz}, {and} \bibinfo{person}{Carlo Ratti}.} \bibinfo{year}{2018}\natexlab{}.
\newblock \showarticletitle{Addressing the minimum fleet problem in on-demand urban mobility}.
\newblock \bibinfo{journal}{\emph{Nature}} \bibinfo{volume}{557}, \bibinfo{number}{7706} (\bibinfo{year}{2018}), \bibinfo{pages}{534--538}.
\newblock


\bibitem[Wang et~al\mbox{.}(2020)]%
        {wang2020data}
\bibfield{author}{\bibinfo{person}{Chao Wang}, \bibinfo{person}{Yi Hou}, {and} \bibinfo{person}{Matthew Barth}.} \bibinfo{year}{2020}\natexlab{}.
\newblock \showarticletitle{Data-driven multi-step demand prediction for ride-hailing services using convolutional neural network}. In \bibinfo{booktitle}{\emph{Advances in Computer Vision: Proceedings of the 2019 Computer Vision Conference (CVC), Volume 2 1}}. Springer, \bibinfo{pages}{11--22}.
\newblock


\bibitem[Wang et~al\mbox{.}(2023)]%
        {wang2023optimization}
\bibfield{author}{\bibinfo{person}{Dujuan Wang}, \bibinfo{person}{Qi Wang}, \bibinfo{person}{Yunqiang Yin}, {and} \bibinfo{person}{TCE Cheng}.} \bibinfo{year}{2023}\natexlab{}.
\newblock \showarticletitle{Optimization of ride-sharing with passenger transfer via deep reinforcement learning}.
\newblock \bibinfo{journal}{\emph{Transportation Research Part E: Logistics and Transportation Review}}  \bibinfo{volume}{172} (\bibinfo{year}{2023}), \bibinfo{pages}{103080}.
\newblock


\bibitem[Wenzel et~al\mbox{.}(2019)]%
        {wenzel2019travel}
\bibfield{author}{\bibinfo{person}{Tom Wenzel}, \bibinfo{person}{Clement Rames}, \bibinfo{person}{Eleftheria Kontou}, {and} \bibinfo{person}{Alejandro Henao}.} \bibinfo{year}{2019}\natexlab{}.
\newblock \showarticletitle{Travel and energy implications of ridesourcing service in {A}ustin, {T}exas}.
\newblock \bibinfo{journal}{\emph{Transportation Research Part D: Transport and Environment}}  \bibinfo{volume}{70} (\bibinfo{year}{2019}), \bibinfo{pages}{18--34}.
\newblock


\bibitem[Xu et~al\mbox{.}(2017)]%
        {xu2017real}
\bibfield{author}{\bibinfo{person}{Jun Xu}, \bibinfo{person}{Rouhollah Rahmatizadeh}, \bibinfo{person}{Ladislau B{\"o}l{\"o}ni}, {and} \bibinfo{person}{Damla Turgut}.} \bibinfo{year}{2017}\natexlab{}.
\newblock \showarticletitle{Real-time prediction of taxi demand using recurrent neural networks}.
\newblock \bibinfo{journal}{\emph{IEEE Transactions on Intelligent Transportation Systems}} \bibinfo{volume}{19}, \bibinfo{number}{8} (\bibinfo{year}{2017}), \bibinfo{pages}{2572--2581}.
\newblock


\bibitem[Xu et~al\mbox{.}(2023)]%
        {xu2023ride}
\bibfield{author}{\bibinfo{person}{Wenjie Xu}, \bibinfo{person}{Qingyu Liu}, \bibinfo{person}{Minghua Chen}, {and} \bibinfo{person}{Haibo Zeng}.} \bibinfo{year}{2023}\natexlab{}.
\newblock \showarticletitle{Ride the Tide of Traffic Conditions: Opportunistic Driving Improves Energy Efficiency of Timely Truck Transportation}.
\newblock \bibinfo{journal}{\emph{IEEE Transactions on Intelligent Transportation Systems}} (\bibinfo{year}{2023}).
\newblock


\bibitem[Zha et~al\mbox{.}(2016)]%
        {zha2016economic}
\bibfield{author}{\bibinfo{person}{Liteng Zha}, \bibinfo{person}{Yafeng Yin}, {and} \bibinfo{person}{Hai Yang}.} \bibinfo{year}{2016}\natexlab{}.
\newblock \showarticletitle{Economic analysis of ride-sourcing markets}.
\newblock \bibinfo{journal}{\emph{Transportation Research Part C: Emerging Technologies}}  \bibinfo{volume}{71} (\bibinfo{year}{2016}), \bibinfo{pages}{249--266}.
\newblock


\end{thebibliography}
